\documentclass[aps,prb,twocolumn,superscriptaddress]{revtex4}
\usepackage{graphicx}
\usepackage{bm}
\usepackage{epsfig} 
\usepackage{amsfonts}
\usepackage{amsmath}

\newcommand*{\be}{\begin{equation}}
\newcommand*{\ee}{\end{equation}}
\newcommand*{\bea}{\begin{eqnarray}}
\newcommand*{\eea}{\end{eqnarray}}

\newcommand*{\sd}{^{\dagger}}

\newcommand{\up}{\uparrow}
\newcommand{\down}{\downarrow}

\def\ket#1{\left| #1\right\rangle}
\def\p{\partial}

\begin{document}


\title{Doped carrier formulation and mean-field theory of the $tt't''J$ model}

\author{Tiago C. Ribeiro}
\affiliation{Department of Physics, University of California, Berkeley, California 94720, USA}
\affiliation{Material Sciences Division, Lawrence Berkeley National Laboratory, Berkeley, California 94720, USA}
\author{Xiao-Gang Wen}
\affiliation{Department of Physics, Massachusetts Institute of Technology, Cambridge, Massachusetts 02139, USA}

\date{\today}

\begin{abstract}
In the generalized-$tJ$ model 
the effect of the large local Coulomb repulsion is
accounted for by restricting the Hilbert space to states with at most one 
electron per site.
In this case the electronic system can be viewed in terms of holes hopping
in a lattice of correlated spins, where holes are the carriers doped
into the half-filled Mott insulator.
To explicitly capture the interplay between the hole dynamics and
local spin correlations we derive a new formulation of the generalized-$tJ$ 
model where doped carrier operators are used instead of the original electron
operators.
This ``doped carrier'' formulation provides a new starting point to address 
doped spin systems and we use it to develop a new, fully fermionic, 
mean-field description of doped Mott insulators. 
This mean-field approach reveals a new mechanism for superconductivity, 
namely spinon-dopon mixing, and we apply it to the $tt't''J$ model
as of interest to high-temperature superconductors.
In particular, we use model parameters borrowed from band
calculations and from fitting ARPES data to obtain a mean-field phase diagram
that reproduces semi-quantitatively that of hole and electron doped cuprates.
The mean-field approach hereby presented accounts for the 
local antiferromagnetic and $d$-wave superconducting correlations which, we
show, provide a rational for the role of $t'$ and $t''$ in strengthening
superconductivity as expected by experiments and other theoretical approaches.
As we discuss how $t$, $t'$ and $t''$ affect the phase diagram,
we also comment on possible scenarios to understand 
the differences between as-grown and oxygen reduced electron doped samples.
\end{abstract}


\maketitle

\section{\label{sec:intro}Introduction}

High-temperature superconducting (SC) cuprates are layered materials where
the copper-oxide planes are separated by several elements whose chemistry
controls the density of carriers in the CuO$_2$ layers.  This density
determines the electronic nature of copper-oxide planes, where the renowned
unconventional cuprate phenomenology is believed to take place.  Due to the
localized character of $3d$ orbitals, copper valence electrons feel a large
Coulomb interaction which drives strong correlations. \cite{A8796} The largest
in-plane energy scale is the local Coulomb repulsion $U$, measured to be
approximately $2$eV, \cite{KB9897} thus motivating the use of the
generalized-$tJ$ model in the cuprate context. \cite{A8796,D9463,LN0445}
Interestingly, upon inclusion of parameters $t'$ and $t''$ consistent with
electronic structure calculations \cite{PD0103} the generalized-$tJ$ model
reproduces many spectral features observed in both hole and electron doped
cuprates. \cite{TM0017,T0417} 

Following its aforementioned relevance, 
in this paper we consider the two-dimensional $tt't''J$ Hamiltonian
\begin{align}
H_{tJ} & = J \sum_{\langle ij \rangle \in NN}
\left( \bm{S}_i.\bm{S}_j -\frac{1}{4} \mathcal{P}n_i n_j\mathcal{P} \right) - 
\notag \\
& \quad - \sum_{\langle ij \rangle} t_{ij} \mathcal{P} 
\left( c_{i}\sd c_{j} + c_{j}\sd c_{i}\right) \mathcal{P}
\label{eq:Htj}
\end{align}
where $t_{ij}=t,t',t''$ for first, second and third nearest neighbor (NN) 
sites respectively.
$c_i\sd = [c_{i,\up}\sd c_{i,\down}\sd]$ is the electron 
creation spinor operator,
$n_i = c_i\sd c_i$ and $\bm{S}_i = c_i\sd \bm{\sigma} c_i$ are the 
electron number and spin operators and $\bm{\sigma}$ are the Pauli matrices. 
The operator $\mathcal{P}$ projects out states with on-site double
electron occupancy and, therefore,
the $tt't''J$ model 
Hilbert space consists of states where every site has either a spin-1/2 
or a vacancy.
Consequently, 
when the number of electrons equals the number of lattice sites, \textit{i.e.}
at half-filling, there exists exactly one electron on each site
and the system is a Mott insulator which can be described in terms of 
spin variables alone.
In the doped Mott insulator case, however, the model Hamiltonian also 
describes electron hopping, which is highly constrained since
electrons can only move onto a vacant site.
This fact is captured by the use of projected electron 
operators $\mathcal{P} c_{i,\sigma}\sd \mathcal{P}$
that do not obey the canonical fermionic anti-commutation relations
and, thus, constitute a major hurdle to handle the $tt't''J$ model 
analytically.
For that reason, the physics of the generalized-$tJ$ model has been 
addressed by a variety of numerical techniques, which include the exact 
diagonalization of small systems, 
\cite{T0417,TM9496,GV9466,MH9545,KW9845,NO9890,R0402} 
the self-consistent Born approximation, 
\cite{RV8893,MH9117,LM9225,NV9576,XW9653,BC9614}
the cellular dynamical mean-field (MF) approximation, \cite{KK0614}
quantum Monte Carlo, \cite{DM9582,MH9545,PH9544,DZ0016,GE0036}
Green function Monte Carlo \cite{DN9428,DN9510} 
and variational Monte Carlo methods. \cite{G8831,SM0202,SL0402,PR0404}
These studies provide an overall consistent picture that guides
the effort to develop analytical approaches to the 
generalized-$tJ$ model, which is the main focus of the current work,
and we refer to them throughout the paper.

Different analytical techniques and approximations have been developed
to address doped Mott insulators in terms of bosonic and/or fermionic 
operators amenable to a large repertoire of many-body physics tools.
The simplest one, known as the Gutzwiller approximation, 
\cite{G6359,ZG8836} trades the projection operators in 
$\mathcal{P} c_{i}\sd c_{j} \mathcal{P}$ 
by a numerical renormalization factor $g_t = \tfrac{2x}{1+x}$ that 
vanishes in the half-filling limit.
The extensively used slave-particle techniques 
\cite{B7675,C8435,BZ8773,BA8880,AZ8845,KL8842,LN9221,WL9603,LN9803,RV8893}
rather decouple the projected electron operator 
$\mathcal{P} c_{i,\sigma}\sd \mathcal{P}$ into a fermion and a boson which 
describe chargeless spin-1/2 excitations of the spin background
and spinless charge $e$ excitations that keep track of the vacant sites.
These various approaches provide distinct ways to handle the 
projected electron operator $\mathcal{P} c_{i,\sigma}\sd \mathcal{P}$ which,
we remark, significantly differs from the bare electron operator $c_i\sd$
close to half-filling. \cite{HFILLING}

In this paper, we present and explicitly derive a different approach to
doped Mott insulators that circumvents the use of projected electron 
operators.
This approach was first introduced in Ref. \onlinecite{RW0501}
and recasts the $tt't''J$ model Hamiltonian \eqref{eq:Htj} in terms 
of spin variables and projected doped hole operators which carry the
charge and spin degrees of freedom introduced in the system upon doping.
To make the difference between both approaches more concrete, note
that while projected electron operators are used to fill the sea
of interacting electrons starting from the empty vacuum, projected doped
hole operators are used to describe the sea of interacting carriers
doped into the half-filled spin system. 

One should bear in mind that the physical properties of holes in 
doped Mott insulators differ from those of holes in conventional band 
insulators.
The distinction is clear in the infinite on-site Coulomb repulsion limit,
\textit{i.e.} in the generalized-$tJ$ model regime, in which case the 
insertion of a hole on a certain site $i$ amounts to the removal of a 
lattice spin.
This leaves a vacant site, that carries a unit charge $e$ when compared to 
the remaining sites,
and changes the total spin component along the $z$-direction by 
$\pm \tfrac{1}{2}$. \cite{POLARON}
Since the vacancy is a spin singlet entity, the extra 
spin $\pm \tfrac{1}{2}$ introduced upon doping is not on the vacant
site and is carried by the surrounding spin background instead.
Hence, in doped Mott insulators, the object that carries the same quantum 
numbers as doped holes, namely charge $e$ and spin $\pm \tfrac{1}{2}$, is
not an on-site entity. \cite{BAND_I}
We remark that in the doped Mott insulator literature the term ``hole'' 
is often used with different meanings. 
Sometimes it refers to the charge-$e$ spin-0 object on the vacant 
site from which a spin was removed.
We reserve the term ``vacancy'' for such an on-site object which differs 
from the above described entity which carries both non-zero charge and spin 
quantum numbers.
We use the term ``doped hole'' to allude to this latter non-local 
entity.
In fact, and since the $tt't''J$ model can be used to describe both the 
lower Hubbard band or the upper Hubbard band in the large Coulomb 
repulsion limit, in this paper we generally use the term ``doped carrier'' 
to mean ``doped hole'' or ``doped electron'' depending on whether we refer 
to the hole or electron doped case.

It follows from the above argument that a doped carrier in doped
Mott insulators is a composite object that involves both the vacancy
and its surrounding spins and,
therefore, its properties are related to those of the background
spin correlations.
Since at half-filling the generalized-$tJ$ model reduces to the Heisenberg
Hamiltonian, which displays antiferromagnetic (AF) order, \cite{M9101}
it is natural to think of doped carriers in terms of a vacancy encircled
by a staggered spin configuration.
As emphasized in Ref. \onlinecite{DN9510}, this picture is valid even in 
the absence of long-range AF order and, in fact, it only requires
that most of the doped carrier spin-1/2 is concentrated in a region
whose linear size is smaller than the AF correlation length.
Interestingly, quantum Monte Carlo calculations for the $U/t=8$
Hubbard model \cite{DM9582} find signatures of short-range AF correlations 
around the vacancy up to the hole density $x=0.25$.
Angle-resolved photoemission spectroscopy (ARPES) experiments
also show that the high energy hump present in undoped samples,
which disperses in accordance with the $tt't''J$ model single hole 
dispersion, \cite{XW9653,BC9614,KW9845,TM0017} is present all the way 
into the overdoped regime. \cite{DS0373}
Hence, both theory and experiments support that short-range AF
correlations persist throughout a vast range of the high-T$_c$ phase 
diagram and, thus, suggest that the above local picture of a doped
carrier holds as we move away from half-filling.

It is important to understand how superconductivity arises out
of the above doped carrier objects.
One possibility is that doped carriers form a gas of
fermionic quasiparticles \cite{DN9428,NO9890} which 
form local bound states in the $d$-wave channel due to AF correlations. 
\cite{DN9510}
Alternatively, it has been proposed \cite{A8796} that doped carriers
induce spin liquid correlations which do not frustrate the hopping of 
charge carriers (unlike AF correlations) and which 
ultimately lead to SC order.
ARPES data on the cuprates identifies two different dispersions --
at low energy the characteristic $d$-wave SC linear dispersion crosses the 
Fermi level close to $(\tfrac{\pi}{2},\tfrac{\pi}{2})$ while at higher
energy the dispersion inherited from the undoped limit persists. 
\cite{RS0301,SR0402,KS0318,YZ0301}
This experimental evidence supports the coexistence of both \textit{local} 
AF and $d$-wave spin singlet correlations and, thus, favors the
second scenario above.
The coexistence of two distinct types of spin correlations also receives
support from exact diagonalization calculations of the $tt't''J$ model.
\cite{R0402}

Since the formulation of the $tt't''J$ model hereby presented 
introduces doped carrier operators to address doped Mott insulators 
close to half-filling we dub it the ``doped carrier'' formulation of 
the $tt't''J$ model.  
A number of advantages stem from using doped carrier rather than the 
original particle operators.
For instance, close to half-filling the doped carrier density is small 
and so we are free of the no-double-occupancy constraint problem for doped 
carriers.
In addition, as we show in this paper, in the ``doped carrier'' framework 
the hopping sector of the $tt't''J$ model explicitly describes the interplay 
between the doped carrier dynamics and different local spin correlations,  
specifically the aforementioned coexisting local AF and $d$-wave SC 
correlations.
Consequently, this approach provides a powerful framework to understand 
the single-particle dynamics in doped Mott insulators,
as attested by Refs. \onlinecite{RW0501,RW_arpes,RW0531} which show
that it can be used to reproduce a large spectrum of 
ARPES and tunneling spectroscopy data on the cuprates.

In Sec. \ref{sec:formulation} we present the detailed derivation of the 
``doped carrier'' formulation of the $tt't''J$ model.
In particular, we define the aforementioned projected doped carrier 
operators, which we call ``dopon'' operators, as on-site fermionic operators 
with spin-1/2 and unit electric charge.
Dopons act on an Hilbert space larger than the $tt't''J$ model 
Hilbert space and thus they are distinct from electron operators.
Only after the contribution from all unphysical states 
is left out do dopons describe the physical doped 
carriers, which are the non-local entities consisting of the vacancy 
and the extra spin $\pm \tfrac{1}{2}$ carried by the encircling spins.

The ``doped carrier'' framework constitutes a new starting point to address
doped spin systems and
in Sec. \ref{sec:mf} we construct a new, fully fermionic, MF 
theory of doped Mott insulators, which we call the ``doped carrier'' MF 
theory.
In this MF approach we represent the lattice spin variables, 
which provide the half-filled background on top of which doped carriers
are added, in terms of fermionic spin-1/2 chargeless excitations known 
as spinons.
Hence, we describe the low energy dynamics of the $tt't''J$
model in terms of spinons and dopons.
Since the latter carry the same quantum numbers as electrons, the ``doped
carrier'' formulation can account for low lying electron-like 
quasiparticles which form Fermi arcs as observed in high
temperature superconductors. \cite{RW0501,RW_arpes,ND9857}
This state of affairs is to be contrasted with that stemming from
the slave-boson approach to the generalized-$tJ$ model.
\cite{BZ8773,BA8880,AZ8845,KL8842,LN9221,WL9603,LN9803} 
In the latter approach the low energy dynamics is described by spinons and 
holons (spinless charge-$e$ bosons) and, thus, it
stresses the physics of spin-charge separation.
Despite the differences, it is possible to relate the ``doped carrier''
and the slave-boson formulations of the generalized-$tJ$ model and,
in Appendix \ref{app:slave_boson}, we show that holons 
are the singlet bound states of spinons and dopons.

In Sec. \ref{sec:pd} we use model parameters extracted both from
ARPES data and from electronic structure calculations concerning 
the copper-oxide layers of high-temperature superconductors to discuss
the MF phase diagrams of interest to hole and electron doped 
cuprates.
The relevance of the hereby proposed approach is supported by the
semi-quantitative agreement between our results and the experimental
phase diagram for this family of materials.
The ``doped carrier'' MF theory explicitly accounts
for how local AF and $d$-wave SC correlations correlations affect and are 
affected by the local doped carrier dynamics and, thus, 
we are able to discuss the role played by the 
hopping parameters $t$, $t'$ and $t''$ in determining the MF phase diagram.
These and other results underlie our conclusions
(Sec. \ref{sec:conclusion}).

\section{\label{sec:formulation}
Doped carrier formulation of the $tt't''J$ model}

\subsection{\label{subsec:enlarged}Enlarged Hilbert space}

The $tt't''J$ model Hamiltonian \eqref{eq:Htj} is written in terms
of projected electron operators $\mathcal{P} c_{i,\sigma}\sd \mathcal{P}$
and $\mathcal{P} c_{i,\sigma} \mathcal{P}$ which rule out doubly 
occupied sites and, thus, the $tt't''J$ model on-site Hilbert space for 
any site $i$ is
\be
\mathcal{H}_i = \{\ket{\uparrow}_i,\ket{\downarrow}_i,\ket{0}_i\}
\label{eq:Hilb}
\ee
The states in \eqref{eq:Hilb} include the spin-up, spin-down and vacancy
states respectively.

In this section, we introduce a different, though equivalent, formulation
of the $tt't''J$ model.
In the usual formulation any wave-function can be written by acting
with the projected electron creation operators 
$\mathcal{P} c_{i,\sigma}\sd \mathcal{P}$ on top of an empty background.
Close to half-filling these operators substantially differ from
the bare electron creation operator and we propose an alternative
framework where the background on top of which carriers are created is a 
lattice of spin-1/2 objects, which we call the ``lattice spins''. 
We then consider fermionic spin-1/2 objects with unit charge,
which we call ``dopons'', that move on top of the lattice spin background.
Dopons have the same spin and electric charge quantum numbers as the 
carriers doped in the system and are introduced to describe these doped 
carriers.
In such a description there is one lattice spin in every site
whether or not this site corresponds to a physical vacancy.
In addition, there exists one, and only one, dopon in every physically
vacant site.
However, a vacant site is a spinless object while dopons carry spin-1/2.
Therefore, to accommodate the presence of both lattice spins and dopons 
we must consider an enlarged on-site Hilbert space which for any
site $i$ is
\be
\mathcal{H}_i^{enl} = \{ \ket{\up\!\!0}_i, \ket{\down\!\!0}_i,
\ket{\up \down}_i, \ket{\down \up}_i, 
\ket{\up \up}_i, \ket{\down \down}_i \}
\label{eq:enl_Hilb}
\ee
States in \eqref{eq:enl_Hilb} are denoted by $\ket{\alpha \, a}$,
where $\alpha=\up,\down$ labels the up and down states of lattice spins 
and $a=0,\up,\down$ labels the three dopon on-site states, namely the
no dopon, spin-up dopon and spin-down dopon states.
To act on these states we introduce the lattice spin operator 
$\widetilde{\bm{S}}_i$ and the fermionic dopon creation spinor operator 
$d_i\sd = [d_{i,\up}\sd d_{i,\down}\sd]$, which are such that
$\widetilde{S}_i^z \ket{\sigma a}_i = s_{\sigma} \frac{1}{2}\ket{\sigma a}_i$
[where $s_{\sigma} = (+)$ for $\sigma = \up$ and 
$s_{\sigma} = (-)$ for $\sigma = \down$] and 
$d_{i,\sigma}\sd \ket{\alpha \, 0}_i = \ket{\alpha \sigma}$.
Since in the enlarged Hilbert space \eqref{eq:enl_Hilb} there exist no 
states with two dopons on the same site we also introduce the 
projection operator 
$\widetilde{\mathcal{P}} = \prod_i (1-d_{i,\up}\sd d_{i,\up} 
d_{i,\down}\sd d_{i,\down})$ 
which enforces the no-double-occupancy constraint for dopons.

In order to write physical operators, like the projected electron 
operators or the $tt''t''J$ model Hamiltonian \eqref{eq:Htj}, in terms 
of lattice spin operators $\widetilde{\bm{S}}_i$ and projected dopon operators 
$\widetilde{\mathcal{P}} d_{i,\sigma}\sd \widetilde{\mathcal{P}}$
and $\widetilde{\mathcal{P}} d_{i,\sigma} \widetilde{\mathcal{P}}$ we
define the following mapping from states in the enlarged on-site Hilbert 
space \eqref{eq:enl_Hilb} onto the physical $tt't''J$ model on-site
Hilbert space \eqref{eq:Hilb}
\be
\ket{\up}_i \leftrightarrow \ket{\up\!\!0}_i \ ; \ \
\ket{\down}_i \leftrightarrow \ket{\down\!\!0}_i \ ; \ \
\ket{0}_i \leftrightarrow 
\frac{\ket{\up \down}_i-\ket{\down \up}_i}{\sqrt{2}}
\label{eq:physical_enl_map}
\ee
The on-site triplet states in $\mathcal{H}_i^{enl}$, namely
$\frac{\ket{\up \down}_i + \ket{\down \up}_i}{\sqrt{2}}$, 
$\ket{\up \up}_i$ and $\ket{\down \down}_i$  are unphysical as they do 
not map onto any state pertaining to the $tt't''J$ model on-site Hilbert 
space.
Therefore, these latter states must be left out when writing down
the physical wave-function or when defining how physical operators act 
on the enlarged Hilbert space.

\subsection{\label{subsec:electron}Electron operator in 
enlarged Hilbert space}
The mapping rules \eqref{eq:physical_enl_map} can be used to define
the spin-up electron creation operator $\tilde{c}_{i,\up}\sd$ that acts 
on the on-site enlarged Hilbert space and whose matrix elements on the 
physical sector of \eqref{eq:enl_Hilb} match those of
$\mathcal{P} c_{i,\up}\sd \mathcal{P}$ on \eqref{eq:Hilb}, namely
\be
\mathcal{P} c_{i,\up}\sd \mathcal{P} \left\{
\ket{\up}_i; \ket{\down}_i; \ket{0}_i \right\} \ = \ \left\{
0; 0; \ket{\up}_i \right\} 
\label{eq:electron_definition}
\ee
If we further require $\tilde{c}_{i,\up}\sd$ 
to vanish when it acts on unphysical states we obtain the relations
\begin{gather}
\tilde{c}_{i,\up}\sd \frac{\ket{\up \down}_i-\ket{\down \up}_i}{\sqrt{2}} = 
\ket{\up\!\!0}_i \notag \\
\tilde{c}_{i,\up}\sd \ket{\up\!\!0}_i = 
\tilde{c}_{i,\up}\sd \ket{\down\!\!0}_i = 0 \notag \\
\tilde{c}_{i,\up}\sd \left(\ket{\up \down}_i + \ket{\down \up}_i\right) =
\tilde{c}_{i,\up}\sd \ket{\up \up}_i = 
\tilde{c}_{i,\up}\sd \ket{\down \down}_i = 0
\label{eq:electron_up_creation}
\end{gather}
It is a trivial matter to recognize that the defining relations 
\eqref{eq:electron_up_creation} are satisfied by the operator
\be
\tilde{c}_{i,\up}\sd = \frac{1}{\sqrt{2}} \widetilde{\mathcal{P}} \left[
\left( \frac{1}{2} + \widetilde{S}_i^z \right) d_{i,\down} -
\widetilde{S}_i^+ d_{i,\up} \right] \widetilde{\mathcal{P}}
\label{eq:electron_up_dagger_enl}
\ee
where $\widetilde{S}_i^{\pm} = \widetilde{S}_i^x \pm i \widetilde{S}_i^y$.

The spin-down electron creation operator can be determined along the
same lines or, alternatively, by considering 
an angle $\pi$ rotation around the $y$-axis, which transforms 
\begin{align}
\tilde{c}_{i,\up}\sd \longrightarrow -\tilde{c}_{i,\down}\sd \ ; \ \
d_{i,\up} &\longrightarrow -d_{i,\down} \ ; \ \
d_{i,\down} \longrightarrow d_{i,\up} \notag \\
\widetilde{S}_i^z \longrightarrow -\widetilde{S}_i^z \ ; \ \
\widetilde{S}_i^+ &\longrightarrow -\widetilde{S}_i^- \ ; \ \
\widetilde{S}_i^- \longrightarrow -\widetilde{S}_i^+
\end{align}

It then follows that the electron operators \cite{ED,E_OPERATOR}
that act on the on-site enlarged Hilbert space are
\be
\tilde{c}_{i,\sigma}\sd = s_{\sigma} \frac{1}{\sqrt{2}}\widetilde{\mathcal{P}} 
\left[\left( \frac{1}{2} + s_{\sigma} \widetilde{S}_i^z \right) 
d_{i,-\sigma} - \widetilde{S}_i^{s_{\sigma}} d_{i,\sigma} \right] 
\widetilde{\mathcal{P}}
\label{eq:electron_dagger_enl}
\ee

\subsection{\label{subsec:tj_enlarged}$tt't''J$ model Hamiltonian in 
enlarged Hilbert space}
We now recast the $tt't''J$ model Hamiltonian \eqref{eq:Htj}, which 
is a function of the projected electron operators
$\mathcal{P} c_{i,\sigma}\sd \mathcal{P}$ and 
$\mathcal{P} c_{i,\sigma} \mathcal{P}$, in terms of 
lattice spin and projected dopon operators as dictated by Expression
\eqref{eq:electron_dagger_enl}.
The resulting $tt't''J$ model Hamiltonian in the enlarged Hilbert space 
\be
H_{tJ}^{enl} = H_{enl}^{J} + H_{enl}^{t}
\label{eq:Htj_enl}
\ee
is the sum of the Heisenberg ($H_{enl}^{J}$) and the hopping 
($H_{enl}^{t}$) terms.
First we use the equalities
\begin{gather}
\widetilde{\mathcal{P}} d_{i,\sigma} \widetilde{\mathcal{P}} d_{i,-\sigma}\sd 
\widetilde{\mathcal{P}} =
\widetilde{\mathcal{P}} ( d_{i,\up} \widetilde{\mathcal{P}} d_{i,\up}\sd -
d_{i,\down} \widetilde{\mathcal{P}} d_{i,\down}\sd ) \widetilde{\mathcal{P}} = 
0  \notag \\
\widetilde{\mathcal{P}} ( d_{i,\up} \widetilde{\mathcal{P}} d_{i,\up}\sd +
d_{i,\down} \widetilde{\mathcal{P}} d_{i,\down}\sd ) \widetilde{\mathcal{P}} = 
2 \widetilde{\mathcal{P}} (1-d_i\sd d_i) \widetilde{\mathcal{P}}
\end{gather}
to replace the operators 
$\bm{S}_i = (\mathcal{P}c_i\sd\mathcal{P}) \bm{\sigma} 
(\mathcal{P}c_i\mathcal{P})$ and 
$\mathcal{P} n_i \mathcal{P} = (\mathcal{P}c_i\sd\mathcal{P}) 
(\mathcal{P}c_i\mathcal{P})$ in the first term of \eqref{eq:Htj} by 
$\widetilde{\bm{S}}_i \widetilde{\mathcal{P}}
(1-d_i\sd d_i) \widetilde{\mathcal{P}}$ and 
$\widetilde{\mathcal{P}} (1-d_i\sd d_i) \widetilde{\mathcal{P}}$
respectively.
As a result, the Heisenberg interaction in terms of lattice spin and
projected dopon operators is
\be
H_{enl}^{J} = J \!\! \sum_{\langle ij \rangle \in NN} \!\!
\left(\widetilde{\bm{S}}_i.\widetilde{\bm{S}}_j - \frac{1}{4}\right) \, 
\widetilde{\mathcal{P}}
\left( 1 - d_i\sd d_i \right) \left( 1 - d_j\sd d_j \right)
\widetilde{\mathcal{P}}
\label{eq:heisenberg_enl}
\ee
We obtain the hopping term in the enlarged Hilbert space upon directly 
replacing $\mathcal{P} c_{i,\sigma}\sd \mathcal{P}$ and 
$\mathcal{P} c_{i,\sigma} \mathcal{P}$ in the second term of 
\eqref{eq:Htj} by $\tilde{c}_{i,\sigma}\sd$ and $\tilde{c}_{i,\sigma}$, 
which leads to
\bea
H_{enl}^{t} &=& \sum_{\langle ij \rangle} \frac{t_{ij}}{2}
\widetilde{\mathcal{P}} \left[ \left( d_i\sd \bm{\sigma} d_j \right) . \left( 
i \widetilde{\bm{S}}_i \times \widetilde{\bm{S}}_j - 
\frac{\widetilde{\bm{S}}_i+\widetilde{\bm{S}}_j}{2} \right)
+ \right. \notag \\
&& \left. \quad \quad \quad \quad \quad + \frac{1}{4} d_i\sd d_j + d_i\sd d_j 
\widetilde{\bm{S}}_i . \widetilde{\bm{S}}_j + h.c. \right] 
\widetilde{\mathcal{P}}
\label{eq:hop_enl}
\eea

The hopping term in the $tt't''J$ model Hamiltonian 
\eqref{eq:Htj} connects a state with a vacancy on site $j$ and a spin on 
site $i$ to that with an equal spin state on site $j$ and a vacancy on 
site $i$.
This is a two site process that leaves the remaining sites unaltered
and, schematically, it amounts to
\begin{gather}
\ket{0;\up} \ \rightarrow \ \ket{\up;0} \notag \\
\ket{0;\down} \ \rightarrow \ \ket{\down;0}
\label{eq:hop_2site}
\end {gather}
where the notation $\ket{j;i}$ is used to represent the states on sites
$j$ and $i$.
Using the corresponding two site notation for the enlarged Hilbert space,
namely $\ket{\alpha_j a_j; \alpha_i a_i}$,
where $\alpha_i,\alpha_j=\up,\down$ and $a_i,a_j=0,\up,\down$, 
and making use of the mapping rules \eqref{eq:physical_enl_map}, 
the hopping processes in the ``doped carrier'' framework that
correspond to \eqref{eq:hop_2site} are
\begin{gather}
\ket{\frac{\up \down - \down \up}{\sqrt{2}};\up\!\!0} \ \rightarrow \ 
\ket{\up\!\!0;\frac{\up \down - \down \up}{\sqrt{2}}} \notag \\
\ket{\frac{\up \down - \down \up}{\sqrt{2}};\down\!\!0} \ \rightarrow \ 
\ket{\down\!\!0;\frac{\up \down - \down \up}{\sqrt{2}}}
\label{eq:hop_2site_enl}
\end{gather}
It can be explicitly shown that the hopping term \eqref{eq:hop_enl} is 
such that its only non-vanishing matrix elements are those that describe 
the processes in \eqref{eq:hop_2site_enl}.
Hence, only local singlet states hop between 
different lattice sites whereas the unphysical
local triplet states are localized and have no kinetic energy. 
Therefore, \textit{the dynamics described by $H_{enl}^{t}$ effectively
implements the local singlet constraint}, which leaves out the unphysical
states in the enlarged Hilbert space \eqref{eq:enl_Hilb}.

We emphasize that the $tt't''J$ Hamiltonian in the enlarged Hilbert space 
\eqref{eq:Htj_enl} equals $H_{tJ}$ in the physical Hilbert space.
In addition, it does not connect the physical and the unphysical sectors 
of the enlarged Hilbert space.
Therefore, the ``doped carrier'' formulation of the $tt't''J$ model, 
as defined by \eqref{eq:Htj_enl}, \eqref{eq:heisenberg_enl} 
and \eqref{eq:hop_enl}, is equivalent to the original ``particle'' 
formulation encoded in \eqref{eq:Htj}.
Interestingly, \textit{it provides a different starting point 
to deal with doped spin models}.

In the low doping regime the dopon density
$x = \tfrac{1}{N}\sum_i \widetilde{\mathcal{P}} d_i\sd d_i 
\widetilde{\mathcal{P}}$, where $N$ is the number of sites, is small and 
the no-double-occupancy constraint for dopons is safely relaxed.
Hence, below we drop all the projection operators $\widetilde{\mathcal{P}}$.
We thus propose that \textit{the dramatic effect of the projection operators
$\mathcal{P}$ in the ``particle'' formulation of the $tt't''J$ model
\eqref{eq:Htj} is captured by the dopon-spin interaction
in the hopping hamiltonian \eqref{eq:hop_enl}},
which \textit{explicitly accounts for the role of local spin correlations on 
the hole dynamics}.
In the remainder of the paper we derive and discuss a MF theory 
that describes this interaction.

\section{\label{sec:mf}Doped carrier mean-field theory of the $tt't''J$ model}

To derive a MF theory of the $tt't''J$ model ``doped carrier'' 
formulation we recur to the fermionic representation
of lattice spins 
$\widetilde{\bm{S}}_i = \tfrac{1}{2} f_i\sd \bm{\sigma} f_i$, 
where $f_i\sd = [f_{i,\up}\sd f_{i,\down}\sd]$ is the spinon creation spinor
operator. \cite{A6505}
The Hamiltonian $H^{enl}_{tJ}$ is then the sum of terms 
with multiple fermionic operators which can be decoupled upon
the introduction of appropriate fermionic averages, as presented in 
what follows.
The resulting MF Hamiltonian $H_{tJ}^{MF}$ is quadratic in the operators 
$f\sd$, $f$, $d\sd$ and $d$, and describes the hopping, pairing and 
mixing of spinons and dopons. 
We remark that, in contrast to slave-particle approaches which split
the electron operator into a bosonic and a fermionic excitations,
\cite{RV8893,LN9221,WL9603}
this MF Hamiltonian is purely fermionic.

\subsection{\label{subsec:mf_heisenberg}Heisenberg term}
We first consider the spin exchange interaction \eqref{eq:heisenberg_enl}
which upon replacing the operator
$\widetilde{\mathcal{P}} (1 - d_i\sd d_i) \widetilde{\mathcal{P}}$
by its average value $(1-x)$ reduces to the Heisenberg 
Hamiltonian
\be 
\tilde{J} \sum_{\langle ij \rangle \in NN} 
\left( \widetilde{\bm{S}}_i.\widetilde{\bm{S}}_j - \frac{1}{4} \right)
\label{eq:renorm_heisenberg}
\ee
with the renormalized exchange constant 
$\tilde{J} = (1-x)^2 J$.

In the enlarged Hilbert space \eqref{eq:enl_Hilb} there is always 
one lattice spin $\widetilde{\bm{S}}_i$ per site, even in the presence
of finite doping.
Therefore, in the fermionic representation of $\widetilde{\bm{S}}_i$
the $SU(2)$ projection constraint enforcing $f_i\sd f_i = 1$ must be 
implemented by adding the term 
$\sum_i \bm{a}_{0,i}. \left(\psi_i\sd \bm{\sigma} \psi_i\right)$
to \eqref{eq:Htj_enl}, where $\bm{a}_{0,i}$ are local Lagrangian 
multipliers. \cite{WL9603,LN9803,W0213}
Here we use the Nambu notation for spinon operators, namely
$\psi_i\sd = [\psi_{i,1}\sd \psi_{i,2}\sd] = [f_{i,\up}\sd f_{i,\down}]$,
in terms of which lattice spin operators can be recast as 
$\widetilde{\bm{S}}_i . \bm{\sigma} = \tfrac{1}{2} (\Psi_i\sd \Psi_i - I)$
where \cite{AZ8845}
\be
\Psi_i = \left[ \begin{array}{cc} 
\psi_{i,1} & \psi_{i,2}\sd \\
\psi_{i,2} & -\psi_{i,1}\sd \end{array}\right]
\label{eq:spinon_matrix}
\ee
and $I$ is the identity matrix.
It then follows that, in the fermionic representation, the Hamiltonian term 
\eqref{eq:renorm_heisenberg} reduces to
\be
-\frac{\tilde{J}}{8} \sum_{\langle ij \rangle \in NN}
Tr\left[\widehat{U}_{i,j} \widehat{U}_{j,i}\right] + 
\sum_{i} \bm{a}_{0,i}. \left(\psi_i\sd \bm{\sigma} \psi_i\right)
\label{eq:heisenberg_2}
\ee
with $\widehat{U}_{i,j} = - \Psi_i \Psi_j\sd$.

The quartic fermionic terms in \eqref{eq:heisenberg_2}
can be decoupled by means of the Hartree-Fock-Bogoliubov 
approximation thus leading to the MF Heisenberg term \cite{WL9603,LN9803}
\begin{align}
H_{MF}^J &= \frac{3\tilde{J}}{16} \sum_{\langle ij \rangle \in NN}
Tr\left[U_{i,j}U_{j,i}\right] + 
\bm{a}_0.\left(\sum_i \psi_i\sd \bm{\sigma} \psi_i \right) -
\notag \\
& \quad - \frac{3\tilde{J}}{8} \sum_{\langle ij \rangle \in NN} 
\left( \psi_i\sd U_{i,j} \psi_j + h.c. \right)
\label{eq:heisenberg_MF}
\end{align}
where $U_{i,j} = \langle \widehat{U}_{i,j} \rangle$ are the singlet bond
MF order parameters
(in Sec. \ref{subsec:mf_af} the above MF decoupling is extended to include
the formation of local staggered moments as well).
Also note that at the MF level the local constraint is relaxed and 
$\bm{a}_0$ is taken to be site independent.
The best non-symmetry breaking MF parameters to describe the paramagnetic
spin liquid state of the Heisenberg model correspond to the ansatz 
\be
U_{i,i+\hat{x}} = \chi \sigma_z + \Delta \sigma_x \ ; \
U_{i,i+\hat{y}} = \chi \sigma_z - \Delta \sigma_x \ ; \
\bm{a}_0 = a_0 \sigma_z
\label{eq:MF_dwave}
\ee
which describes spinons paired in the $d$-wave channel and
whose properties have been studied in the context of the slave-boson 
approach. \cite{KL8842,WL9603,LN9803,LN0445}

\subsection{\label{subsec:mf_hop}Hopping term}

We now consider the hopping Hamiltonian \eqref{eq:hop_enl},
which describes the hopping of holes on the top of a spin background
with strong local AF correlations.
It is well understood that, under such circumstances, the hole
dispersion is renormalized by spin fluctuations. 
\cite{KL8980,RV8893,D9463,DN9428}
As it is further elaborated in Sec. \ref{subsec:tp_tpp}, to capture the 
effect of this renormalization at the MF level we replace the bare $t$, $t'$ 
and $t''$ by effective hopping parameters $t_1$, $t_2$ and $t_3$.

The hopping term \eqref{eq:hop_enl} involves both spinon operators
\eqref{eq:spinon_matrix} and dopon operators
\be
D_i = \left[ \begin{array}{cc} 
d_{i,\up} & d_{i,\down} \\
d_{i,\down}\sd & -d_{i,\up}\sd \end{array}\right]
\label{eq:dopon_matrix}
\ee
and after dropping the projection operators $\widetilde{\mathcal{P}}$ 
it can be recast as
\begin{align}
\sum_{\langle ij \rangle} \frac{t_{ij}}{8} & \left\{
Tr \left[ \left( \Psi_j\sd\Psi_j - I\right) D_j\sd \sigma_z D_i 
\left(\Psi_i\sd\Psi_i - I\right)\right] + \right. \notag \\
& \left. + Tr\left[ D_i \Psi_i\sd\Psi_i D_j\sd\right] +
Tr\left[ D_j \Psi_j\sd\Psi_j D_i\sd\right] + \right. \notag \\
& \left. + Tr\left[ D_j\sd \sigma_z D_i \right] \right\}
\label{eq:hop_su2matrix}
\end{align}
The contribution from \eqref{eq:hop_su2matrix} to the MF Hamiltonian,
namely $H_{MF}^t$, can be decomposed into the MF terms that belong
to the spinon sector ($H_{MF}^{t,spinon}$), the ones that contribute to 
the dopon sector ($H_{MF}^{t,dopon}$) and the ones that mix spinons and 
dopons ($H_{MF}^{t,mix}$), so that 
$H_{MF}^t = H_{MF}^{t,spinon} + H_{MF}^{t,dopon} + H_{MF}^{t,mix}$.
For the sake of clarity, in what follows, we discuss each of the above 
three contributions separately:

\textit{(i)} 
$H_{MF}^{t,spinon}$ arises from decoupling the first term in
\eqref{eq:hop_su2matrix}. 
The decoupling is done by taking the average
\be
\langle \Psi_j D_j\sd \sigma_z D_i \Psi_i\sd \rangle =
\langle \widehat{B}_{j,j}\sd \sigma_z \widehat{B}_{i,i}\rangle \approx 
x \sigma_z \delta_{\langle ij \rangle \in NN}
\label{eq:b_square}
\ee
where we introduce the spinon-dopon singlet pair operator
$\widehat{B}_{j,i} = D_j \Psi_i\sd$.
Then \cite{PREFACTOR}
\be
H_{MF}^{t,spinon} = \sum_{\langle ij \rangle \in NN} \frac{t_1 x}{2} 
\left(\psi_i\sd \sigma_z \psi_j + h.c. \right)
\label{eq:spinon_hop_contribution}
\ee
This term determines the effect of spinon-dopon pairs $\widehat{B}_{j,i}$ 
in the magnitude of the spinon $d$-wave pairing gap.
As we discuss in Appendix \ref{app:slave_boson} the spinon-dopon singlet
pair operator $\widehat{B}_{j,i}$ corresponds to the holon operator in the 
slave-boson formalism and, in fact, a term similar to
\eqref{eq:spinon_hop_contribution} appears in the $SU(2)$ slave-boson MF
Hamiltonian. \cite{RW0301}

\textit{(ii)} 
The MF dopon hopping term $H_{MF}^{t,dopon}$ comes from the fourth 
term in \eqref{eq:hop_su2matrix} and from taking the average
\begin{align}
\langle \left( \Psi_i\sd\Psi_i - I\right) \left( \Psi_j\sd\Psi_j - I\right)
\rangle 
&= 4 \langle \widetilde{\bm{S}}_i.\widetilde{\bm{S}}_j + 
i \bm{\sigma}.\left( \widetilde{\bm{S}}_i \times \widetilde{\bm{S}}_j
\right) \rangle \notag \\
&\approx (-1)^{j_x+j_y-i_x-i_y}
\label{eq:s_square}
\end{align}
in the first term of \eqref{eq:hop_su2matrix}.
As we mention in Sec. \ref{sec:intro}, both numerical 
\cite{DN9510,DM9582,DN9428,MH9545,NO9890,PH9544} and experimental
\cite{DS0373,RS0301,SR0402,KS0318,YZ0301} evidence point to the
persistence of signatures due to local AF correlations around the 
vacancies all the way into the doping regime where cuprates superconduct.
To account for this effect we consider that the spins 
encircling the vacancy in the one-dopon state are in a \textit{local}
N\'eel configuration and, therefore, in \eqref{eq:s_square} we use 
$\langle \widetilde{\bm{S}}_i \times \widetilde{\bm{S}}_j\rangle = 0$
and 
$4 \langle \widetilde{\bm{S}}_i . \widetilde{\bm{S}}_j\rangle = 
(-1)^{j_x+j_y-i_x-i_y}$.
As a result, the contribution from the first and fourth terms in 
\eqref{eq:hop_su2matrix} to the MF dopon sector is
\begin{align}
H_{MF}^{t,dopon} &= \sum_{\langle ij \rangle\in 2^{nd} NN} \!\! \frac{t_2}{4} 
\left( \eta_i\sd \sigma_z d_j + h.c. \right) + \notag \\
& \quad + \sum_{\langle ij \rangle\in 3^{rd} NN} \!\! \frac{t_3}{4} 
\left( \eta_i\sd \sigma_z \eta_j + h.c. \right)
\label{eq:dopon_hop}
\end{align}
where the dopon Nambu operators 
$\eta_i\sd \equiv [\eta_{i,1}\sd \eta_{i,2}\sd ] \equiv 
[d_{i,\up}\sd d_{i,\down}]$ are introduced.
Ideally, the average in \eqref{eq:s_square} should be calculated
self-consistently to reproduce the doping
induced change in the local spin correlations.
In the present MF scheme this is not performed and, instead, we
introduce doping dependent effective hopping parameters $t_2$ and
$t_3$ (see Sec. \ref{subsec:tp_tpp}).

\textit{(iii)} 
The MF term that mixes dopons and spinons, $H_{MF}^{t,mix}$,
captures the interaction between the spin degrees of freedom and the doped
carriers enclosed in the second and third terms of 
\eqref{eq:hop_su2matrix}, which can be recast as
\be
\sum_{\langle ij \rangle} \frac{t_{ij}}{8}
Tr\left[ \widehat{B}_{j,i}\sd \widehat{B}_{i,i} + 
\widehat{B}_{i,j}\sd \widehat{B}_{j,j} \right]
\label{eq:quartic_operator}
\ee
In the Hartree-Fock-Bogoliubov approximation the above expression 
leads to the spinon-dopon mixing term
\begin{align}
H_{MF}^{t,mix} &= - \sum_{i} Tr\left[ B_{i1}\sd B_{i0} \right] - 
\sum_{i} \left[ \eta_i\sd B_{i1} \psi_i + h.c. \right]  
\notag \\
& \ \ \
- \frac{3}{16} \sum_{i,\nu} t_{\nu} \sum_{\hat{u}\in \nu \, NN}
\left( \eta_{i+\hat{u}} \sd B_{i0} \psi_i + h.c. \right)
\label{eq:mf_interaction}
\end{align}
where $B_{i0} = B_{i,i}$ and
$B_{i1} = \tfrac{3}{16} \sum_{\nu} t_{\nu} \sum_{\hat{u}\in \nu \, NN} 
B_{i+\hat{u},i}$.
Here, the mean-fields $B_{j,i} = \langle \widehat{B}_{j,i} \rangle$ are 
introduced.
We also use $\hat{u} = \pm\hat{x},\pm\hat{y}$,
$\hat{u} = \pm\hat{x}\pm\hat{y}$ and $\hat{u} = \pm2\hat{x},\pm2\hat{y}$
for $\nu=1,2,3$ respectively.
Different choices for the mean-fields $B_{i0}$ and $B_{i1}$ may describe 
distinct MF phases.
In what follows we take
\begin{gather}
B_{i0} = - b_0 \sigma_z \ ; \ \
B_{i1} = -b_1 \sigma_z
\label{eq:mf_b_ansatz}
\end{gather}
which, as we show in Appendix \ref{app:slave_boson}, 
describes the $d$-wave SC phase when taken together with the spinon 
$d$-wave ansatz \eqref{eq:MF_dwave}.

In order to clarify the physical picture enclosed in the above
MF scheme, note that dopons correspond to vacancies surrounded by a
staggered spin configuration and, therefore, they describe
quasiparticles in the half-filling limit.
Such a \textit{locally} AF spin background strongly
suppresses coherent doped carrier inter-sublattice hopping
\cite{D9463,KL8980,DN9428} as captured by $H_{MF}^{t,dopon}$ which 
includes dopon hopping processes between $2^{nd}$ and $3^{rd}$ NN sites 
but not between $1^{st}$ NN sites. 
At MF level, the only mechanism for dopons to hop between different 
sublattices is provided by the spinon-dopon mixing term 
\eqref{eq:mf_interaction} which represents the interaction between dopons 
and the lattice spins.
This interaction leads to the formation of spinon-dopon pairs 
$\widehat{B}_{j,i}$, which are spin singlet electrically charged objects 
and, thus, describe vacancies surrounded by spin singlet correlations
that enhance the hopping of charge carriers.
From \eqref{eq:mf_b_ansatz} we have that 
$b_0 = \langle f_i\sd d_i\rangle$ and
$b_1 = \langle\tfrac{3}{16}\sum_{\nu} t_{\nu} \sum_{\hat{u}\in \nu \, NN}
f_i\sd d_{i+\hat{u}}\rangle$ are the local and non-local MF 
parameters that emerge from such spin assisted doped carrier hopping
events and which describe the hybridization of spinons and dopons.
In the rest of the paper we interchangeably refer to the condensation
of the bosonic mean-fields $b_0$ and $b_1$ as the coherent spinon-dopon 
hybridization, mixing or pairing.
As a final remark, note that for $b_1\!\!\neq\!\!0$ the term 
$\sim b_1 d_i\sd f_i$ in  
\eqref{eq:mf_interaction} drives local mixing of spinons and dopons
and leads to a non-zero $b_0$.
Similarly, if $b_0\!\!\neq\!\!0$ the term 
$\sim b_0 d_{i+\hat{u}}\sd f_i$ leads to non-local spinon-dopon mixing and,
thus, to non-zero $b_1$.
Hence, either $b_0$ and $b_1$ are both zero or both non-zero.

\subsection{\label{subsec:mf_ham}Mean-field Hamiltonian}

Putting the terms \eqref{eq:heisenberg_MF}, 
\eqref{eq:spinon_hop_contribution}, \eqref{eq:dopon_hop} and
\eqref{eq:mf_interaction} together leads to the full 
``doped carrier'' MF Hamiltonian $H_{tJ}^{MF} = H_{MF}^J + H_{MF}^t$,
which in momentum space becomes
\begin{align}
H_{tJ}^{MF} &= \sum_{\bm{k}} \left[ 
\begin{array}{cc} \psi_{\bm{k}}\sd & \eta_{\bm{k}}\sd \end{array} 
\right] \left[
\begin{array}{cc} \alpha_{\bm{k}}^z \sigma_z + \alpha_{\bm{k}}^x \sigma_x &
\beta_{\bm{k}} \sigma_z \\ \beta_{\bm{k}} \sigma_z & \gamma_{\bm{k}} \sigma_z
\end{array}
\right] \left[
\begin{array}{c} \psi_{\bm{k}} \\ \eta_{\bm{k}} \end{array}
\right] + \notag \\
& \quad + \frac{3\tilde{J}N}{4} (\chi^2+\Delta^2) - 2Nb_0b_1 - N\mu_d
\label{eq:HMF}
\end{align}
where
\begin{gather}
\alpha_{\bm{k}}^z = -\left(\frac{3\tilde{J}}{4}\chi - 
t_1 x \right) \left(\cos k_x+\cos k_y\right)+a_0 \notag \\
\alpha_{\bm{k}}^x = - \frac{3\tilde{J}}{4} \Delta 
\left(\cos k_x-\cos k_y\right) \notag \\
\begin{split}
\beta_{\bm{k}} &= \frac{3b_0}{8}\left[t_1\left(\cos k_x+\cos k_y\right) + 
2 t_2\cos k_x \cos k_y + \right. \notag \\
& \quad \quad \quad \left.+ t_3 \left(\cos 2k_x+\cos 2k_y\right)\right] + b_1 
\end{split} \notag \\
\gamma_{\bm{k}} = t_2 \cos k_x \cos k_y + \frac{t_3}{2}
\left(\cos 2k_x+\cos 2k_y\right)-\mu_d
\label{eq:HMF_matrix_elements}
\end{gather}
In \eqref{eq:HMF} 
we introduce the dopon chemical potential $\mu_d$ that sets the 
doping level $\bigl<d_i\sd d_i\bigr> = x$.
The explicit form of the above MF Hamiltonian depends on the values of 
$t_1$, $t_2$ and $t_3$, which are determined
phenomenologically in Section \ref{subsec:tp_tpp} by fitting to both 
numerical results and cuprate ARPES data.
The mean-field parameters $\chi$, $\Delta$, $b_0$, and $b_1$
are determined by minimizing the mean-field free-energy and in 
Section \ref{sec:pd} we show they reproduce the cuprate phase diagram.

\subsection{\label{subsec:multiband} Two-band description of
one-band $tt't''J$ model}

Even though the $tt't''J$ model is intrinsically a one-band model, the 
above MF approach contains two different families of spin-1/2 fermions,
namely spinons and dopons, and thus presents a two-band description of the 
same model.
As a result, $H_{tJ}^{MF}$ has a total of four fermionic bands described 
by the eigenenergies
\begin{gather}
\epsilon_{1,\bm{k}}^{\pm}=\pm\sqrt{\rho_{\bm{k}}-\sqrt{\delta_{\bm{k}}}}
\notag \\
\epsilon_{2,\bm{k}}^{\pm}=\pm\sqrt{\rho_{\bm{k}}+\sqrt{\delta_{\bm{k}}}} 
\label{eq:eigenenergies}
\end{gather}
where
\begin{gather}
\delta_{\bm{k}}=\beta_{\bm{k}}^2\left[\left(
\gamma_{\bm{k}}+\alpha_{\bm{k}}^z\right)^2+
\left(\alpha_{\bm{k}}^x\right)^2\right]+\frac{1}{4}\left[\gamma_{\bm{k}}^2-
\left(\alpha_{\bm{k}}^x\right)^2 - \left(\alpha_{\bm{k}}^z\right)^2\right]^2 
\notag \\
\rho_{\bm{k}}=\beta_{\bm{k}}^2 + \frac{1}{2} \left[\gamma_{\bm{k}}^2+
\left(\alpha_{\bm{k}}^x\right)^2+\left(\alpha_{\bm{k}}^z\right)^2\right] 
\label{eq:eigenenergies_terms}
\end{gather}

In the absence of spinon-dopon mixing, \textit{i.e.} when $b_0,b_1=0$,
the bands $\epsilon_{1,\bm{k}}^{\pm}$ and $\epsilon_{2,\bm{k}}^{\pm}$
describe, on the one hand, the spinon $d$-wave dispersion that underlies
the same spin dynamics as obtained by slave-boson theory. 
\cite{RW0201}
In addition, these bands also capture the dispersion of a hole
surrounded by staggered local moments which includes only 
intra-sublattice hopping processes [Expression \eqref{eq:dopon_hop}], 
as appropriate in the one-hole limit of the $tt't''J$ model. 
\cite{D9463,KL8980,DN9428}
Upon the hybridization of spinons and dopons the eigenbands 
$\epsilon_{1,\bm{k}}^{\pm}$ and $\epsilon_{2,\bm{k}}^{\pm}$
differ from the bare spinon and dopon bands by a term of order
$b_0^2,b_1^2,b_0b_1 \sim x$.
In particular, the lowest energy bands $\epsilon_{1,\bm{k}}^{\pm}$
are $d$-wave-like with nodal points along the $(0,0)-(\pm\pi,\pm\pi)$ 
directions and describe electronic excitations that coherently hop
between NN sites.
The highest energy bands $\epsilon_{2,\bm{k}}^{\pm}$ are mostly
derived from the bare dopon bands and, therefore, describe excitations
with reduced NN hopping. \cite{RW0501,RW_arpes}

The reason underlying the above multi-band description of the interplay 
between spin and local charge dynamics stems from the strongly correlated 
nature of the problem.
Quasiparticles in conventional uncorrelated materials correspond to
dressed electrons whose dispersion depends to a small extent on the 
remaining excitations and is largely determined by an effective 
external potential.
In the presence of strong electron interactions, though, the dynamics
of electronic excitations is intimately connected to the surrounding
environment and depends on the various local spin correlations.
Physically, the two-band description provided by the ``doped carrier''
MF theory captures the role played by two such different local spin 
correlations on the hole dynamics.
These are the local staggered moment correlations, which are driven by
the exchange interaction but frustrate NN hole hopping, and the $d$-wave spin 
liquid correlations, which enhance NN hole hopping at the cost of spin 
exchange energy.

We remark that such a multi-band structure agrees with quantum Monte Carlo 
and cellular dynamical MF theory calculations on the two-dimensional 
$tJ$ and Hubbard models. \cite{MH9545,PH9544,DZ0016,GE0036,KK0614} 
In Refs. \onlinecite{DZ0016,GE0036} the two bands below 
the Fermi level were interpreted in terms of two different states, namely:
(i) holes on the top of an otherwise unperturbed spin background and
(ii) holes dressed by spin excitations.
This interpretation is consistent with additional numerical work 
indicating the existence of two relevant spin configurations around the 
vacancy \cite{R0402} and offers support to the above MF formulation.

Finally, we point out that the two-band description of the 
generalized-$tJ$ model ``doped carrier'' formulation resembles that of 
heavy-fermion models: 
dopons and lattice spins in the ``doped carrier'' framework correspond
to conduction electrons and to the spins of $f$-electrons, respectively,
in heavy-fermion systems.
The main difference is that, at low dopings, the $tt't''J$ model 
spin-spin interaction 
is larger than the dopon Fermi energy, while in heavy-fermion models
the spin-spin interaction between $f$-electrons is much smaller than 
the Fermi energy of conduction electrons.
As the doping concentration increases the dopon Fermi energy approaches,
and can even overcome, the interaction energy between lattice spins.
In that case, our approach to the $tt't''J$ model becomes qualitatively 
similar to heavy-fermion models.  
Interestingly, overdoped cuprate samples do behave like heavy-fermion 
systems except for the mass enhancement, which is not as large as for 
typical heavy-fermion compounds.

\subsection{\label{subsec:tp_tpp}Renormalized hopping parameters}

In Sec. \ref{subsec:mf_hop} we mention that at the MF level we resort to 
effective hopping parameters $t_1$, $t_2$ and $t_3$
to account for the renormalization due to spin fluctuations.
In this paper we take the NN hopping parameter to equal its
bare value, \textit{i.e.} $t_1=t$.
However, as we discuss in what follows, the role of local spin 
correlations on the intra-sublattice doped carrier dynamics is quite 
non-trivial and $t_2$ and $t_3$ differ from the corresponding bare 
parameters.

\subsubsection{\label{subsubsec:one_hole}Single hole limit}

The ``doped carrier'' MF theory of the $tt't''J$ model considers the
dilute vacancy limit where we can focus on the local problem of a single 
vacancy surrounded by spins.
In particular, it captures the effect of strong AF correlations around the 
vacancy through the average \eqref{eq:s_square} which determines the dopon 
dispersion $\gamma_{\bm{k}}$. 
This dispersion is controlled by the hopping parameters $t_2$ and $t_3$ and 
does not involve NN hopping processes, in agreement
with $tt't''J$ model single hole problem results. 
\cite{XW9653,BC9614,KW9845,KL8980,TM0017,DN9428,DN9510,MH9117,LM9225}
However, the values of $t_2$ and $t_3$ that fit the above single
hole dispersion differ from the bare $t'$ and $t''$.
This is clearly so when $t'=t''=0$ and, still, charge carriers move 
coherently within the same sublattice. 
\cite{DN9428,MH9117,LM9225}
These intra-sublattice hopping processes, which induce non-zero effective
hopping parameters $t_2$ and $t_3$, result from a spin fluctuation induced 
contribution to charge hopping. \cite{KL8980} 
We remark that this is inherently a quantum contribution that escapes the 
realm of MF theory, which is a semiclassical saddle-point approach,
and therefore we recur to numerical and experimental evidence to set
the values of $t_2$ and $t_3$.

It is well established that for $t'=t''=0$ the single hole dispersion 
has its minimum at $(\pi/2,\pi/2)$ and is quite flat along 
$(0,\pi)-(\pi,0)$. \cite{TM0017,KW9845,DN9428,MH9117,LM9225}
For our purposes, we can simply consider that the hole dispersion is 
completely flat along $(0,\pi)-(\pi,0)$ and, thus, we set $t_2 = 2 t_3 = t_J$ 
when $t'$ and $t''$ vanish.
In this case, numerical calculations show that the dispersion width 
along $(0,0)-(\pi,\pi)$ is $\approx 2J$ \cite{TM0017,D9463} and we take 
$t_J=J$.
The resulting MF effective hopping parameters $t_2=J$ and $t_3=0.5J$ 
compare well with those found by the self-consistent Born approximation 
for $t=0.3J$, namely $t_2=0.87J$ and $t_3=0.62J$, \cite{MH9117} and with 
those determined by the Green function Monte Carlo technique for $t=0.4J$, 
specifically $t_2=0.85J$ and $t_3=0.65J$. \cite{DN9428}

Since non-zero values of $t'$ and $t''$ do not frustrate nor are 
frustrated by local AF correlations, in this case we naively take the 
effective intra-sublattice hopping parameters introduced in 
Sec. \ref{subsec:mf_hop} to be
\begin{gather}
t_2 = t_J + t' \notag \\
t_3 = \frac{t_J}{2} + t''
\label{eq:effective_hop_1hole}
\end{gather}
Based on experiments \cite{WS9564,KW9845} and band theory calculations 
\cite{PD0103} relevant to the cuprates which show that bare intra-sublattice 
hopping parameters are non-zero, \cite{NV9576,GV9466} together with 
numerical calculation results that support $t' \approx -2t''$, 
\cite{XW9653,BC9614,KW9845,TM0017} we set $t' = 2t''$.
The effective hopping parameter choice in \eqref{eq:effective_hop_1hole}
then leads to a dopon dispersion width along $(0,\pi)-(\pi,0)$ equal to 
$2t'$, which given the simple approximation scheme involved is reasonably 
close to numerical results findings, namely that the dispersion width along
$(0,\pi)-(\pi,0)$ is $A t'$ where the coefficient $A$
is somewhere in the range $\tfrac{8}{3} - 4$. \cite{TS0009}

We emphasize that the above comments to the single hole dispersion
in the $tt't''J$ model appear to be relevant to the cuprate materials.
Indeed, a large body of experimental evidence shows that the nodal
dispersion width as measured by ARPES in undoped samples is 
$\approx 2J$, \cite{RS0301,KS0318,SR0402,DS0373,KW9845,TY0403}
where $J$ is independently determined by band calculations \cite{HS9068} 
and from fitting Raman scattering \cite{SF9025} and neutron scattering 
\cite{CH0177} experiments.
In addition, the experimental dispersion along $(0,\pi)-(\pi,0)$
varies for different cuprate families as expected from the above 
mentioned properties of the $tt't''J$ model dispersion and the 
values of $t'$ and $t''$ predicted by band theory. \cite{TY0403,PD0103}
This state of affairs, namely the agreement between $tt't''J$ model
predictions and experimental observations, offers support to
the relevance of this model in the underdoped regime of cuprates.

\subsubsection{\label{subsubsec:nzero_density}Non-zero hole density}

The aforementioned experimental results concern ARPES measurements on 
undoped cuprate samples.
There is, however, experimental evidence that the dispersion present
in these samples persists as a broad high energy hump even away from
half-filling and in the presence of SC long-range order. 
\cite{DS0373,RS0301,SR0402} 
In Refs. \onlinecite{RW0501,RW_arpes} this high energy dispersive feature 
is paralleled to the $\epsilon_{2,\bm{k}}^-$ band in the ``doped carrier''
MF theory, which up to corrections of order $x$ is given by the
bare dopon dispersion $\gamma_{\bm{k}}$ and, thus, by 
$t_2$ and $t_3$.

The above hump energy at $(0,\pi)$ and $(\pi,0)$, also known as the
high energy pseudogap, lowers continuously as the hole concentration
is increased. \cite{DS0373,TY0403}
In order to reproduce this experimental evidence within the ``doped
carrier'' MF approach the effective hopping parameters $t_2$ and $t_3$ 
must be doping dependent.
In particular, note that if $t_2 = 2t_3 = t_J$ the resulting dispersion
is flat along the $(0,\pi)-(\pi,0)$ line.
Therefore, in order to reproduce the decrease of the high energy pseudogap
scale, the $t_2$ and $t_3$ doping dependence must come from the doping
induced renormalization of $t'$ and $t''$.
As a result, in the presence of non-zero hole density we use
\begin{gather}
t_2 = t_J + r(x) t' \notag \\
t_3 = \frac{t_J}{2} + r(x) t''
\label{eq:effective_hop}
\end{gather}
instead of Expression \eqref{eq:effective_hop_1hole}.
Here, $r(x)$ is a renormalization parameter which satisfies $r(0)=1$ and 
that decreases with increasing $x$.

The above doping induced renormalization of $t'$ and $t''$ is suggested
from comparison to experiments.
However, below we argue that such a renormalization is consistent with
theoretical studies of the $tt't''J$ model.
In Sec. \ref{subsec:mf_hop} we show that the dopon dispersion and, thus,
$\epsilon_{2,\bm{k}}^-$ as well, depends on the local spin correlations
that enter the average \eqref{eq:s_square}.
In the present MF approach this average is set by hand and is not
calculated self-consistently.
Therefore, it misses the doping induced changes in the underlying
local spin correlations.
In Ref. \onlinecite{R0402} it is explicitly shown that, in the $tt't''J$ 
model, the spin correlations induced by a hole hopping in a lattice of 
antiferromagnetically correlated spins strongly frustrate $t'$ and $t''$.
Hence, we expect that upon calculating the spin average \eqref{eq:s_square}
self-consistently the above renormalization of intra-sublattice hopping 
processes is properly reproduced.

The just mentioned theoretical results indicate that the renormalization 
coefficient $r(x)$ should decrease with $x$, however, they give no 
information toward its explicit functional dependence.
We thus recur to experimental data which indicates that the high 
energy pseudogap scale vanishes around $x \approx 0.30$, 
\cite{DS0373,TY0403} to chose $r(x)$ to vanish at $x=0.30$.
In addition, we consider $r(x)$ to interpolate linearly between
its $x=0$ and $x=0.30$ values, which yields
$r(x) = \left(1-\tfrac{x}{0.3}\right)$.
We have also considered alternative interpolation schemes (not shown), 
say by changing the exponent of $\left(1-\tfrac{x}{0.3}\right)$ from 
$1$ to $2$, without affecting our general conclusions.

We remark that, even though $t_2$ and $t_3$ are used to control 
the high energy dispersion $\epsilon_{2,\bm{k}}$ in consonance with 
experiments,
there is no such direct experimental input on the low energy band
$\epsilon_{1,\bm{k}}$, and all its properties result from the
theory.
Interestingly, Refs. \onlinecite{RW0501,RW_arpes,RW0531} find that a 
variety of low energy spectral properties associated with the 
$\epsilon_{1,\bm{k}}$ bands are consistent with ARPES and tunneling
experiments on both hole and electron doped cuprates.
We note that the cuprate hole doped (HD) regime can be addressed using 
$t'\approx -2t'' \approx -J$, \cite{PD0103,KW9845} which
within the context of the ``doped carrier'' MF theory reduces to
using the effective hopping parameters
\begin{gather}
t_2^{HD} = J - J\left(1-\frac{x}{0.3}\right) \notag \\
t_3^{HD} = \frac{J}{2} + \frac{J}{2}\left(1-\frac{x}{0.3}\right) 
\label{eq:effective_hop_HD}
\end{gather}
In the electron doped (ED) regime $t'$ and $t''$ change sign 
\cite{T0417,TM9496} and, hence, in this case $t_2$ and $t_3$ become
\begin{gather}
t_2^{ED} = J + J\left(1-\frac{x}{0.3}\right) \notag \\
t_3^{ED} = \frac{J}{2} - \frac{J}{2}\left(1-\frac{x}{0.3}\right)
\label{eq:effective_hop_ED}
\end{gather}

\subsection{\label{subsec:mf_af}Staggered magnetization decoupling channel}

Both theoretical \cite{M9101} and experimental 
\cite{KB9897,TI8954,LL9081,FK0314} evidence support that the above
MF theory Hamiltonian \eqref{eq:HMF}, which assumes a spin liquid 
background, breaks down at and close to half-filling, where long-range 
AF order sets in.
Therefore, here we extend the ``doped carrier'' MF approach in order 
to account for the staggered magnetization decoupling channel.
Specifically, we introduce
\be
m = (-1)^{i_x+i_y}\langle S_i^z \rangle 
\label{eq:staggered_mag_spin}
\ee
and
\be
n = -\frac{(-1)^{i_x+i_y}}{16} \langle \sum_{\nu=2,3} t_{\nu} 
\sum_{\hat{u}\in \nu \,  NN} \eta_i\sd \eta_{i+\hat{u}} + h.c.\rangle
\label{eq:staggered_mag_dopon}
\ee
which are the lattice spin staggered magnetization and the dopon staggered
magnetization respectively.

The contribution from the above decoupling channels adds to \eqref{eq:HMF}
so that we obtain the new MF Hamiltonian which allows for the presence of 
the AF phase
\begin{align}
H_{AF}^{MF} &= H_{tJ}^{MF} + 2 J^*Nm^2 - 4Nmn - \notag \\
& \quad -2\left(J^*m-n\right) \sum_{\bm{k}} \psi_{\bm{k}+(\pi,\pi)}\sd 
\psi_{\bm{k}} - \notag \\
& \quad -2m\sum_{\bm{k}}(\gamma_{\bm{k}}+\mu_d) 
\eta_{\bm{k}+(\pi,\pi)}\sd \eta_{\bm{k}}
\label{eq:HMF_AF}
\end{align}

It is well known that the above MF AF decoupling scheme overestimates
the strength of magnetic moments. 
To effectively include the effect of fluctuations, which decrease
the staggered magnetization, we introduce a renormalized exchange 
constant $J^*=\lambda \tilde{J}$ in the staggered magnetization 
decoupling channel. \cite{BL0202}
The renormalization factor is determined upon fitting the MF 
staggered magnetization \textit{at half-filling} to the quantum Monte-Carlo
estimate $m=0.31$. \cite{S9778}
In the present case, this condition requires $\lambda=0.34$. \cite{LAMBDA}

\section{\label{sec:pd}Doped carrier mean-field phase diagram}

Starting from the MF Hamiltonian \eqref{eq:HMF_AF} the MF phase diagram
can be computed for different values of doping $x$ and temperature $T$ 
by requiring the self-consistency of the MF parameters 
and by determining the value of the Lagrange 
multipliers $\mu_d$ and $a_0$ that enforce the doping density 
($\langle d_i\sd d_i\rangle = x$) and the global $SU(2)$ projection 
($\langle f_i\sd f_i\rangle = 1$) constraints. 
If we ignore states with coexisting AF and SC orders we can separately
consider those cases when both $b_0,b_1 \neq 0$ and those cases when
both $m,n \neq 0$.
The corresponding saddle-point conditions can then be cast analytically,
as shown in Appendix \ref{app:self_cons} for $m,n=0$ and in 
Appendix \ref{app:self_cons_AF} for $b_0,b_1=0$.

\begin{figure}
\begin{center}
\includegraphics[width=0.49\textwidth]{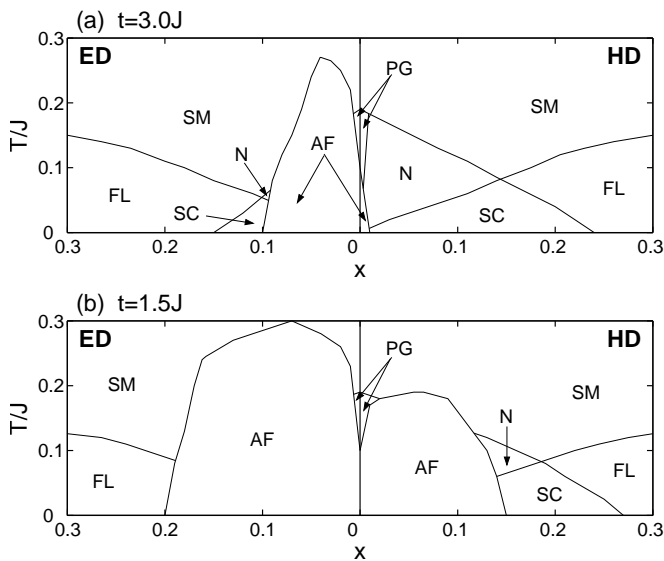}
\caption
{\label{fig:pdiagram_2}
The ``doped carrier'' MF phase diagram
includes the states: AF, $d$-wave SC (dSC), strange metal (SM), 
Fermi liquid (FL), and pseudogap with and without Nernst signal, 
labeled by N and PG respectively.
The hopping parameters (a) $t_1=t=3J$ and (b) $t_1=t=1.5J$ are
used, together with \eqref{eq:effective_hop_HD} and 
\eqref{eq:effective_hop_ED} for the hole doped (HD) and 
electron doped (ED) regimes respectively.
}
\end{center}
\end{figure}

The generic ``doped carrier'' MF theory phase diagram was first 
computed in Ref. \onlinecite{RW0501}.
Here, we show the MF phase diagram for a new set of parameter values, 
namely for NN hopping $t_1=t=3J$ [Fig. \ref{fig:pdiagram_2}(a)] and  
$t_1=t=1.5J$ [Fig. \ref{fig:pdiagram_2}(b)] and for the intra-sublattice 
hopping parameters $t_2$ and $t_3$ in Expressions
\eqref{eq:effective_hop_HD} and \eqref{eq:effective_hop_ED}.
We remark that $t=3J$ together with \eqref{eq:effective_hop_HD} describe
the parameter regime of relevance to HD cuprates and that 
$t=3J$ together with \eqref{eq:effective_hop_ED} address the ED regime.
We consider the $t=1.5J$ case in order to 
illustrate the role of $t$ on
the local energetics (see Sec. \ref{subsec:role_hop}).

Fig. \ref{fig:pdiagram_2} shows that the MF phase diagram
includes six different regions
\cite{REGIONS} which correspond to distinct physical regimes.
These regimes have been discussed within the context of
slave-boson MF theory \cite{LN0445,WL9603,LN9803,PHASES} and, in what
follows, we briefly review their properties:

\textit{(i) Antiferromagnet (AF)} -- 
$\langle b_0 \rangle,\langle b_1 \rangle=0$ and $m,n \neq 0$.
At and close to half-filling the lattice spins form local staggered moments 
and, thus, the low energy spin excitations are spin waves.
Charge carriers move within each sublattice and form small
Fermi pockets whose volume equals the doping level.

\textit{(ii) Strange metal (SM)} -- 
$\langle b_0 \rangle,\langle b_1 \rangle,\Delta,m,n = 0$.
In this spin liquid state the low lying excitations are spinons,
which have an ungaped Fermi surface and, as such, lead to a large 
low energy spin density of states.
Since spinons, which are charge neutral, do not coherently mix with 
dopons, which are charged, the resulting phase is an incoherent metal.

\textit{(iii) Pseudogap metal (PG)} -- 
$\langle |b_0| \rangle,\langle |b_1| \rangle,m,n = 0$ and $\Delta \neq 0$.
When spinons pair up in the $d$-wave channel a gap opens in the uniform
susceptibility in agreement with the observed reduction
of the Knight shift in the pseudogap regime. \cite{CI9777}
Despite the gap, spin correlations at $(\pi,\pi)$ are enhanced
by the gapless $U(1)$ gauge field, \cite{RW0201} as expected in the
underdoped regime close to half-filling.

\textit{(iv) Nernst regime (N)} --
$\langle b_0 \rangle,\langle b_1 \rangle,m,n=0$ and
$\langle |b_0| \rangle, \langle |b_1| \rangle, \Delta \neq 0$.
Below the MF spinon-dopon pairing temperature 
$\langle |b_0| \rangle, \langle |b_1| \rangle \neq 0$ and 
the motion of spinons leads to a backflow in the charged $b_{0,1}$ 
fields so that, effectively, spinons transport electric charge.
Consequently, in the presence of $d$-wave spinon pairing ($\Delta \neq 0$) 
the system displays $d$-wave SC correlations.
Since the magnitude of the spinon-dopon pairing field vanishes toward
half-filling, in the underdoped regime phase fluctuations may prevent
the onset of true long-range SC order, leading to a region in phase
space where $\langle b_0 \rangle,\langle b_1 \rangle = 0$ even though
$\langle |b_0| \rangle, \langle |b_1| \rangle \neq 0$.
In this region, which we call the Nernst region, 
short-range SC fluctuations can be experimentally detected through the 
Nernst effect \cite{XO0086,OW0409,US0201,US0415} or the diamagnetic 
response. \cite{WL0502}

\textit{(v) $d$-wave superconductor (dSC)} -- 
$m,n=0$ and $\langle b_0 \rangle, \langle b_1 \rangle, \Delta \neq 0$.
Below the Kosterlitz-Thouless transition temperature for the above 
mentioned phase fluctuations, the fields $b_0$ and $b_1$ display long-range
phase coherence and the electronic system is a $d$-wave superconductor.

\textit{(vi) Fermi liquid (FL)} -- 
$\Delta,m,n=0$ and $\langle b_0 \rangle, \langle b_1 \rangle \neq 0$.
Since spinons, which are not superfluid ($\Delta = 0$), hybridize with 
dopons they effectively are charged spin-1/2 fermionic excitations with
a large Fermi surface.
Therefore, the electronic system is in the Fermi liquid state, as  
expected from the Ioffe-Larkin sum rule. \cite{IL8988}

Fig. \ref{fig:pdiagram_2}(a) depicts the MF phase diagram in the parameter
regime of interest to the cuprates and shows that
antiferromagnetism is very 
feeble on the HD side leaving room for a large pseudogap region, 
which is mostly covered by the Nernst regime, \cite{NERNST} and for a 
large SC dome.
In the ED case, however, AF order is more robust and covers
a considerable fraction of the SC dome, which is 
smaller than for the HD regime, as well as nearly all
of the Nernst region, in conformity with the lack of experimental 
evidence for a vortex induced Nernst signal in these materials. \cite{BH0320}
In fact, the above MF phase diagram not only
captures the asymmetry between the HD and ED regimes, in agreement
with previous numerical studies, \cite{T0417,TM9496}
but is semi-quantitatively consistent with that of real materials.
Indeed, based on the experimental and numerical input referred to
in Secs. \ref{subsec:tp_tpp} and \ref{subsec:mf_af} we find that:
the SC dome extends up to $x\approx0.24$ on the HD
side while coming to an end at $x\approx0.16$ for the ED
regime;
the maximum T$_c$ is $\sim 100$K in the HD case while it is 
smaller on the ED side.
Furthermore, the renormalized $J^*$, which controls the strength of local 
magnetic moments, is determined by the behavior of the MF theory at 
half-filling and, yet, it correctly predicts that AF order ceases to exist
at a doping level 
which is consistent with experiments on 
both HD and ED compounds. \cite{KB9897,TI8954,LL9081,FK0314}

\subsection{\label{subsec:role_hop}The role of $t$, $t'$ and $t''$}

\begin{figure}
\begin{center}
\includegraphics[width=0.49\textwidth]{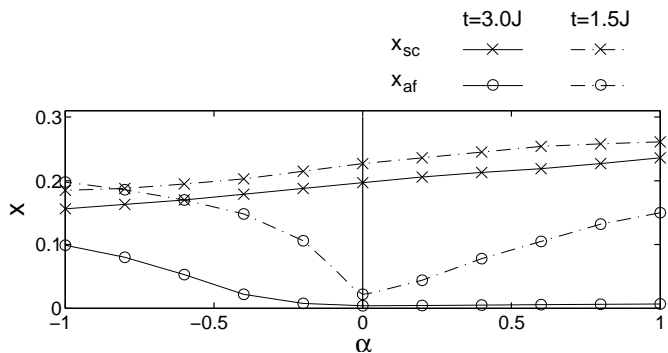}
\caption
{\label{fig:transition_x}
Plot of the doping level $x_{af}$ at which $d$-wave SC order replaces 
AF order (circles) and of the maximum doping level $x_{sc}$ of the 
$d$-wave SC dome (crosses) as a function of $t'=-2t''= -\alpha J$ 
for both $t=3.0J$ (solid line) and $t=1.5J$ (dash-dot line).
Note that $\alpha = 1$ and $\alpha = -1$ reproduce the hole doped 
parameters in \eqref{eq:effective_hop_HD} and the electron doped 
parameters in \eqref{eq:effective_hop_ED} respectively.
}
\end{center}
\end{figure}

The phase diagrams in Fig. \ref{fig:pdiagram_2} show that as we
move away from half-filling the AF phase is initially replaced 
by a state where spins are paired in the $d$-wave singlet channel.
This $d$-wave gapped spin liquid state enhances the doped carrier 
kinetic energy while preserving much of the magnetic exchange
energy. \cite{G8831}
As the doping level increases carrier motion further frustrates the 
exchange energy and eventually closes the $d$-wave gap.
Therefore, the doping evolution of phases is closely connected
to the doped carrier dynamics and,
below, we explore the microscopic picture that underlies how the hopping
parameters $t$, $t'$ and $t''$ affect the robustness of AF and SC 
correlations (see the different phase diagrams in Fig. \ref{fig:pdiagram_2}
as well as how $x_{af}$ and $x_{sc}$, which stand for the doping 
level at which $d$-wave SC order replaces AF order and the maximum 
doping level of the $d$-wave SC dome, change with $t$ and $t' = -2t''$ in 
Fig. \ref{fig:transition_x}).

The combined effect of the $t$, $t'$ and $t''$ terms may enhance
or deplete the hopping between first, second and third NN sites and,
thus, increase or decrease the coupling between doped carriers and
specific surrounding spin correlations.
The particular case of NN hopping is controlled by $t$ and it
frustrates the spin correlations induced by the exchange term in the 
Hamiltonian.
Hence, increasing $t/J$ enfeebles antiferromagnetism and reduces the
$d$-wave spin pairing amplitude $\Delta$, as supported by
the decrease of both $x_{af}$ and $x_{sc}$ in 
Fig. \ref{fig:transition_x} when we go from $t=3J$ to $t=1.5J$.

The hopping between second NN sites is controlled both by $t$ and 
$t'$ --
if the latter parameter is positive then $t$ and $t'$ processes interfere
constructively to enhance second NN hopping, while if $t'<0$
these processes interfere destructively to deplete second
NN hopping.
The same argument applies to third NN hopping if we take $t''$
instead of $t'$ above.
Since both second and third NN hopping do not harm the staggered spin
configuration of the AF state, increasing either $t'$ (as in the
ED side) or $t''$ (as in the HD side) stabilizes AF order and
increases $x_{af}$ (Fig. \ref{fig:transition_x}).
$t'$ in the ED regime is larger than $t''$ in the HD regime and, therefore,
this effect is more prominent in the former case, leading to the
aforementioned asymmetry in the phase diagrams.

Figs. \ref{fig:pdiagram_2} and \ref{fig:transition_x} support that
the values of $t'$ and $t''$ also affect how doped carrier motion couples 
to $d$-wave SC correlations.
To understand this effect note that, in a $d$-wave superconductor, the
condensate induces intra-sublattice hopping processes where the
amplitude for a hole to hop between second NN sites is negative
while the amplitude for a hole to hop between third NN sites is positive.
These processes, and thus SC order as well, are frustrated when $t'=-2t''>0$.
A different way to picture the above argument is to note that when
$t'=-2t''>0$ the doped carrier dispersion is gaped at 
$(\tfrac{\pi}{2},\tfrac{\pi}{2})$, in which case, a vacancy hopping
in the presence of local AF correlations frustrates the 
spinon $d_{x^2-y^2}$-wave dispersion and weakens superconductivity.
Therefore, $x_{sc}$ decreases when $t'$ and $t''$ vary between the HD and 
ED regimes (Fig. \ref{fig:transition_x}).
The same applies to the highest SC T$_c$, in agreement with
experiments and other theoretical approaches. 
Indeed, band theory calculations together with experimental data 
support that the maximum T$_c$ for various cuprate families 
increases with $-t'/t$. \cite{PD0103}
ARPES results also suggest the correlation between the high energy pseudogap 
scale, which is controlled by $t'$ and $t''$, and the maximum T$_c$.
\cite{TY0403}
In addition, variational Monte Carlo calculations further substantiate
the above role of $t'$ in determining the robustness of SC correlations.
\cite{SL0402}

The previous digression on the roles played by $t$, $t'$ and $t''$ in the 
interplay between spin and charge degrees of freedom may be of interest
to understand the striking differences between as-grown and oxygen
reduced electron doped cuprate compounds.
The former samples are not SC and display long-range AF order up to 
$x \approx 0.20$. \cite{MV0402}
After the oxygen reduction process, which removes about $1\%$ of the
oxygen atoms in these materials, AF order is destroyed at 
$x \approx 0.10-0.14$ and, at higher doping values, the samples superconduct.
\cite{TU8997,LL9081,FK0314}
Within the present context, we propose this sharp change follows
the alteration of the effective in-plane parameters  $t$, $t'$ and $t''$.
Since both $t'$ and $t''$ depend on the chemical composition outside
the copper-oxide layers, \cite{PD0103}
if such a composition changes in a way that the magnitude of $t'$ and $t''$
decreases, the phase-space volume of the AF phase is reduced while SC 
correlations are enhanced.
Alternatively, if the oxygen reduction process acts on the copper-oxide
planes in such a manner that $t$ effectively decreases, then
$\tfrac{t}{J} \sim \tfrac{U}{t}$ increases, 
which favors superconductivity over antiferromagnetism.
The latter scenario receives support from experimental evidence
for the removal of oxygen atoms from the copper-oxide planes
under the reduction process in both PCCO \cite{RR0411} and
NCCO. \cite{RR0413}

\section{\label{sec:conclusion}Summary}

In this paper we explicitly derive a recently introduced \cite{RW0501}
formulation of the $tt't''J$ model in terms of projected dopon and 
spin operators instead of projected electron and spin operators.
Since dopons describe the carriers doped in the half-filled system, 
we name it the ``doped carrier'' formulation of the $tt't''J$ model.
Close to half-filling the doped carrier density is small and, thus,
we circumvent the ``no-double-occupancy'' constraint.
In particular, we propose that the effect of the projection operators 
$\mathcal{P}$ in the usual formulation of the $tt't''J$ model
\eqref{eq:Htj} is captured by the interaction between doped carriers and 
lattice spins in \eqref{eq:hop_enl}.
This interaction explicitly accounts for the interplay between local
spin correlations and the hole dynamics in doped Mott insulators.

The $tt't''J$ model Hamiltonian in the enlarged Hilbert space  
[Expression \eqref{eq:Htj_enl}] provides a new starting point
to deal with doped spin models, which we pursue to develop a new,
fully fermionic, MF theory of doped Mott insulators.
The resulting ``doped carrier'' MF theory is constructed to address
the low doping and low temperature 
regime of the $tt't''J$ model, and properly accounts for the
frustration of NN hopping due to the strong local AF correlations
present in such a limit.
Since a hole hopping in an antiferromagneticaly correlated
spin background induces new local spin correlations that strongly 
renormalize $t'$ and $t''$, \cite{R0402} in \eqref{eq:effective_hop}
we introduce a phenomenological doping dependent renormalization 
factor $r(x)$.
Specifically, we choose $r(x)$ so that the high energy MF dispersion
$\epsilon_{2,\bm{k}}^-$ reproduces the evolution of
the high energy pseudogap scale at $(0,\pi)$ observed in ARPES 
experiments. \cite{DS0373,TY0403}
Remarkably, using $t$, $t'$ and $t''$ motivated by band theory 
\cite{HS9068,PD0103} and with the aforementioned little experimental 
input the ``doped carrier'' MF theory leads to a semi-\textit{quantitative} 
correct phase diagram for both HD and ED cuprates 
[Fig. \ref{fig:pdiagram_2}(a)].
In particular, in the HD case a large SC dome and extended pseudogap
regime are obtained, while superconductivity is much weaker on the 
ED side where it is partly overtaken by the robust AF phase.

In the low doping limit of the generalized-$tJ$ model it is meaningful, 
and we believe useful as well, to think of vacancies encircled by local 
moments.
The ``doped carrier'' approach then provides a framework to address the
hole dynamics in the presence of various local spin correlations.
In the hereby developed MF theory, an effective multi-band description
captures the effect of two different local correlations, namely 
staggered moment and $d$-wave singlet bond correlations:
the vacancy in the one-dopon state is encircled by staggered local
moments which inhibit NN hopping;
upon spinon-dopon mixing the vacancy changes the surrounding 
spin background and gains kinetic energy, which is the driving
force for the $d$-wave SC state at low doping.
The above two-band description receives support from more 
rigorous calculations on the two-dimensional $tJ$ and Hubbard models,
\cite{MH9545,PH9544,DZ0016,GE0036,R0402,KK0614} and is consistent
with the various spectral features observed by ARPES experiments.
\cite{RS0301,KS0318,YZ0301,SR0402}
We remark that in the present approach superconductivity arises due
to the change imposed on spin correlations by the motion of doped 
carriers.
This process is captured by the spinon-dopon mixing, which can be 
viewed as a new mechanism for superconductivity.
We believe this mechanism to be relevant for the experimentally
obtained low values of $J/t$.
In the large $J/t$ regime a distinct mechanism, namely the minimization
of the number of missing AF links, may lead doped carriers to form
local pairs, as proposed by the ``antiferromagnetic-van Hove model''. 
\cite{DN9510}

Since the ``doped carrier'' approach addresses the interplay between 
the doped carrier dynamics and different background correlations, we 
can discuss how the strength of the latter 
depends on the hopping parameters.
Interestingly, we find that short-range AF correlations can enhance or 
deplete $d$-wave SC correlations depending on the sign of $t'$ and $t''$.
Specifically, if these hopping parameters favor a gap in the single
hole $tt't''J$ model dispersion at $(\tfrac{\pi}{2},\tfrac{\pi}{2})$, 
as in the ED regime, 
the vacancies when in the presence of local AF configurations frustrate 
the $d$-wave SC gap and, thus, superconductivity as well. 
The opposite effect occurs if, instead, $t'$ and $t''$ induce a gap at 
$(0,\pi)$ in the single hole $tt't''J$ model dispersion.
Hence, the interplay between short-range AF and $d$-wave SC correlations 
provides a \textit{microscopic} rational for the role of $t'$ and $t''$
in strengthening superconductivity as expected by experiments and other
theoretical approaches. \cite{PD0103,TY0403,SL0402}
In this context, note that the ``doped carrier'' MF theory 
not only accounts for the hole/electron doped asymmetry but
also offers possible scenarios for the difference between
phase diagrams of as-grown and oxygen reduced
electron doped samples. \cite{MV0402,TU8997,LL9081,FK0314}

The ``doped carrier'' formulation provides a MF theory to describe
\textit{superconductors with strong local AF correlations 
due to a proximate Mott insulating state}.  
The signature of such correlations is explicit in a
variety of cuprate experimental data that deviates from the pure BCS behavior.
In particular, Refs. \onlinecite{RW0501,RW_arpes,RW0531} show that several
non-trivial features of the electron spectral function and of the tunneling
conductance spectrum of cuprates are reproduced by the herein presented MF
approach.

\begin{acknowledgments}
The authors acknowledge several discussions with P.A. Lee.
This work was supported by the Funda\c c\~ao 
Calouste Gulbenkian Grant No. 58119 (Portugal), 
by the NSF Grant No. DMR--04--33632,
NSF-MRSEC Grant No. DMR--02--13282 and NFSC Grant No. 10228408.
TCR was also supported by the LDRD program of LBNL under DOE
$\#$DE-AC02-05CH11231.

\end{acknowledgments}

\appendix
\section{\label{app:slave_boson}Relation to slave-boson approach}
In the present paper a new formulation of the $tt't''J$ model is
used as a starting point to develop a new MF theory of doped Mott
insulators.
This ``doped carrier'' MF theory, which to the authors' best knowledge
differs from other MF theories in the literature, bears some relations 
to the slave-boson MF approach to the same model. \cite{SLAVE_BOSON}
Both approximations recur to the fermionic representation of spin operators
and, thus, in the undoped limit both MF approaches are equivalent.
However, these MF theories deal with doped carriers in distinct manners
since they introduce different operators to account for charged degrees of 
freedom in doped systems -- the slave-boson formulation introduces
the spinless charged bosonic holon operator 
$h_i\sd = [h_{i,1}\sd h_{i,2}\sd]$ while the ``doped carrier'' formulation
introduces the spin-1/2 charged fermionic dopon operator $d_i\sd$.
In this appendix we clarify the relation between holon and dopon operators
and, thus, the relation between the extensively used slave-boson approach
and the new ``doped carrier'' formalism.

Within slave-boson theory, the projected electron operators, which are 
the building blocks of all other physical operators, can be written in 
terms of holons and spinons as
\begin{align}
\mathcal{P} c_{i,\up}\sd \mathcal{P} &= \frac{1}{\sqrt{2}} \left(
h_{i,1} f_{i,\up}\sd + h_{i,2}f_{i,\down} \right) \notag \\
\mathcal{P} c_{i,\down}\sd \mathcal{P} &= \frac{1}{\sqrt{2}} \left(
h_{i,1} f_{i,\down}\sd - h_{i,2}f_{i,\up} \right)
\label{eq:electron_sb_2}
\end{align}
as long as we constrain ourselves to the physical Hilbert space defined
by $(\psi_i\sd \bm{\sigma} \psi_i + h_i\sd \bm{\sigma} h_i) = 0$.
\cite{WL9603,LN9803}
This formulation introduces an $SU(2)$ gauge structure
since physical operators are invariant under the local transformation
$\psi_i \rightarrow W_i \psi_i$ and $h_i \rightarrow W_i h_i$, where
$W_i$ is any $SU(2)$ matrix.
Hence, both spinons and holons carry an $SU(2)$ gauge charge
in addition to the physical spin and electric charge quantum numbers.

Interestingly, in the ``doped carrier'' formalism dopons and spinons
can form singlet pairs, as captured by the operators 
$\widehat{B}_{i,j}$ introduced in Sec. \ref{subsec:mf_hop}.
These are electrically charged spinless bosonic fields which also carry
the above $SU(2)$ gauge charge and,
therefore, they have the same quantum numbers as holon operators
in the slave-boson framework.
To make the relation between holons and spinon-dopon pairs explicit
it is convenient to rewrite the projected electron operators in terms 
of dopons and spinons.
We use $\widetilde{S}_i^z = \tfrac{1}{2} \left(
f_{i,\up}\sd f_{i,\up} - f_{i,\down}\sd f_{i,\down} \right)$,
$\widetilde{S}_i^+ = f_{i,\up}\sd f_{i,\down}$ and
$\widetilde{S}_i^- = f_{i,\down}\sd f_{i,\up}$ to express the operators
in \eqref{eq:electron_dagger_enl} as
\begin{align}
\mathcal{P} c_{i,\up}\sd \mathcal{P} &= \frac{1}{\sqrt{2}} f_{i,\up}\sd
\widetilde{\mathcal{P}} \left( f_{i,\up}d_{i,\down} - f_{i,\down}d_{i,\up}
\right) \widetilde{\mathcal{P}} \notag \\
\mathcal{P} c_{i,\down}\sd \mathcal{P} &= \frac{1}{\sqrt{2}} f_{i,\down}\sd
\widetilde{\mathcal{P}} \left( f_{i,\up}d_{i,\down} - f_{i,\down}d_{i,\up}
\right) \widetilde{\mathcal{P}}
\label{eq:electron_dc_1}
\end{align}
Since the above equality only holds for $f_i\sd f_i = 1$ it can be
recast as
\begin{align}
\mathcal{P} c_{i,\up}\sd \mathcal{P} &= \frac{1}{\sqrt{2}} f_{i,\down}
\widetilde{\mathcal{P}} \left( f_{i,\up}\sd d_{i,\up} + 
f_{i,\down}\sd d_{i,\down} \right) \widetilde{\mathcal{P}} \notag \\
\mathcal{P} c_{i,\down}\sd \mathcal{P} &= -\frac{1}{\sqrt{2}} f_{i,\up}
\widetilde{\mathcal{P}} \left( f_{i,\up}\sd d_{i,\up} + 
f_{i,\down}\sd d_{i,\down} \right) \widetilde{\mathcal{P}}
\label{eq:electron_dc_2}
\end{align}
Using the slave-boson formulation language, in the physical Hilbert
space there can be at most one holon per site.
Say it happens to be a $h_{i,1}$ holon. 
Then, the projected electron operators in \eqref{eq:electron_sb_2}
resemble those in \eqref{eq:electron_dc_1} with the $h_{i,1}$ holon
replaced by the spinon-dopon pair 
$\left( f_{i,\up}d_{i,\down} - f_{i,\down}d_{i,\up} \right)$.
If, instead, there exists a $h_{i,2}$ holon on site $i$, 
the projected electron operators in \eqref{eq:electron_sb_2}
resemble those in \eqref{eq:electron_dc_2} with the $h_{i,2}$ holon
replaced by the spinon-dopon pair 
$\left( f_{i,\up}\sd d_{i,\up} + f_{i,\down}\sd d_{i,\down} \right)$.
In this sense, the $h_{i,1}$ and $h_{i,2}$ holon operators are related to 
specific singlet pairs of spinons and dopons.

The above correspondence between holons and spinon-dopon pairs
can be used to compare MF phases in the slave-boson and 
``doped carrier'' approaches.
Specifically, for a given spinon state, the physical symmetries of 
a phase described by a certain pattern of holon condensation 
$\langle h_i\sd\rangle=[\langle h_{i,1}\sd \rangle \langle h_{i,2}\sd \rangle]$
in the slave-boson formulation are the same as those of a state where 
spinon-dopon pairing in the ``doped carrier'' formulation yields
\be
B_{i0} = \left[ \begin{array}{cc}
-\langle h_{i,2} \rangle & \langle h_{i,1} \rangle \\
\langle h_{i,1}\sd \rangle & \langle h_{i,2}\sd \rangle
\end{array} \right]
\ee
This result shows that holon condensation in the slave-boson formalism
leads to the same phases as the spinon-dopon pairing transition
observed in the ``doped carrier'' approach.
In particular, in the presence of $d$-wave paired spinons 
described by ansatz \eqref{eq:MF_dwave}, the $d$-wave SC state
obtained upon the condensation of holons 
$\langle h_i\sd \rangle = [0 \ h_0]$
is the same as the one that occurs in the presence of the spinon-dopon
pairing described in \eqref{eq:mf_b_ansatz}.

To understand the connection between holons and spinon-dopon pairs it is 
useful to consider the local picture of these objects.
The dopon is an entity which carries charge and spin and is formed by
the vacancy plus a neighboring spin.
More precisely, the dopon spin is carried by the staggered local moments
that surround the vacancy in the one-dopon state.
Since these local moments frustrate doped carrier hopping
the spin state encircling the vacancy in the one-dopon state is
modified to optimize the doped carrier kinetic energy.
In the ``doped carrier'' MF theory this interaction between doped carriers
and the surrounding spins is captured by the $\sim d\sd f f\sd d$ 
term in \eqref{eq:hop_su2matrix} which can be recast as $\sim b f\sd d$.
This term drives the decay of the dopon into
a spinless spinon-dopon pair and a chargeless spinon.
Physically, this process means that the spin background forms a singlet
with the dopon spin thus altering the spin configuration around the vacancy.
The vacancy is then encircled by a local spin singlet configuration, which 
corresponds to the local picture of a holon.
The doped carrier spin-1/2 is absorbed by the spin background in the form
of a spinon excitation.

The above picture suggests that the holon is a composite object and that
the ``doped carrier'' approach captures its internal structure.
Such an internal structure should then be apparent in the electronic
spectral properties.
In Ref. \onlinecite{RW_arpes} these properties are discussed at length
and the holon internal structure is argued to be reflected in the
momentum space anisotropy  
(also known as the nodal-antinodal dichotomy \cite{ZY0401})
of the electron spectral function.

\section{\label{app:self_cons}Paramagnetic mean-field self-consistency
equations}

Below we write the set of self-consistent conditions for the 
paramagnetic MF Hamiltonian \eqref{eq:HMF} which are determined
by the saddle point equations 
\be
\frac{\p F}{\p \chi} = \frac{\p F}{\p \Delta} = \frac{\p F}{\p b_0} =
\frac{\p F}{\p b_1} = \frac{\p F}{\p \mu_d} = \frac{\p F}{\p a_0} = 0
\ee
where $F$ is the MF free-energy of the fermionic system described by 
\eqref{eq:HMF}, namely
\begin{align}
\frac{F}{N} &= \frac{3\tilde{J}}{4}\left(\chi^2+\Delta^2\right) - 
2b_0b_1 - \mu_d(1-x) - \notag \\
& \quad - \frac{T}{N} \sum_{\bm{k}} 
\ln\left[1+\cosh\left(\frac{\epsilon_{1,\bm{k}}^+}{T}\right)\right]
\left[1+\cosh\left(\frac{\epsilon_{2,\bm{k}}^+}{T}\right)\right]
\label{eq:free_energy}
\end{align}

The resulting self-consistent equations for the mean-fields $\chi$, $\Delta$,
$b_0$ and $b_1$, as well as the doping concentration and $SU(2)$ projection
constraints, are: 
\begin{widetext}
\be
\chi = -\frac{1}{4N} \sum_{\bm{k}} \left(\cos k_x + \cos k_y\right)
\left\{ \alpha_{\bm{k}}^z A_{\bm{k}} - \frac{2\gamma_{\bm{k}}\beta_{\bm{k}}^2+
\alpha_{\bm{k}}^z\left[2 \beta_{\bm{k}}^2+\left(\alpha_{\bm{k}}^x\right)^2+
\left(\alpha_{\bm{k}}^z\right)^2-\gamma_{\bm{k}}^2\right]}
{2\sqrt{\delta_{\bm{k}}}} B_{\bm{k}}\right\} 
\label{eq:selfconsistent_eq_first}
\ee
\be
\Delta =-\frac{1}{4N} \sum_{\bm{k}} \alpha_{\bm{k}}^x
\left(\cos k_x - \cos k_y\right) \left\{A_{\bm{k}} -
\frac{2\beta_{\bm{k}}^2+\left(\alpha_{\bm{k}}^x\right)^2+\left(
\alpha_{\bm{k}}^z\right)^2-\gamma_{\bm{k}}^2}{2\sqrt{\delta_{\bm{k}}}}
B_{\bm{k}}\right\} 
\ee
\begin{align}
b_1 = - \frac{3}{16N} \sum_{\bm{k}} \beta_{\bm{k}} & \left[
t_1\left(\cos k_x + \cos k_y\right) + 2 t_2 \cos k_x \cos k_y +
t_3 \left(\cos 2k_x + \cos 2k_y\right)\right] \times \notag \\
& \times \left\{A_{\bm{k}} - 
\frac{\left(\gamma_{\bm{k}}+\alpha_{\bm{k}}^z\right)^2+\left(\alpha_{\bm{k}}^x
\right)^2}{2\sqrt{\delta_{\bm{k}}}} B_{\bm{k}}\right\} 
\end{align}
\be
b_0 = -\frac{1}{2N} \sum_{\bm{k}}\beta_{\bm{k}} \left\{ A_{\bm{k}} -
\frac{\left(\gamma_{\bm{k}}+\alpha_{\bm{k}}^z\right)^2+\left(\alpha_{\bm{k}}^x
\right)^2}{2\sqrt{\delta_{\bm{k}}}} B_{\bm{k}}\right\} 
\ee
\be
x = 1-\frac{1}{2N} \sum_{\bm{k}} \left\{ \gamma_{\bm{k}}A_{\bm{k}} -
\frac{2\alpha_{\bm{k}}^z\beta_{\bm{k}}^2+\gamma_{\bm{k}}\left[
2\beta_{\bm{k}}^2+\gamma_{\bm{k}}^2-\left(\alpha_{\bm{k}}^x\right)^2-
\left(\alpha_{\bm{k}}^z\right)^2\right]}{2\sqrt{\delta_{\bm{k}}}} B_{\bm{k}}
\right\} 
\ee
\be
0 = \frac{1}{2N} \sum_{\bm{k}} \left\{ \alpha_{\bm{k}}^z A_{\bm{k}} - 
\frac{2\gamma_{\bm{k}}\beta_{\bm{k}}^2+
\alpha_{\bm{k}}^z\left[2 \beta_{\bm{k}}^2+\left(\alpha_{\bm{k}}^x\right)^2+
\left(\alpha_{\bm{k}}^z\right)^2-\gamma_{\bm{k}}^2\right]}
{2\sqrt{\delta_{\bm{k}}}} B_{\bm{k}}\right\}
\label{eq:selfconsistent_eq_last}
\ee
\end{widetext}
where we introduce
\be
A_{\bm{k}} = \frac{\sinh\left(\frac{\epsilon_{1,\bm{k}}^+}{T}\right)}
{\epsilon_{1,\bm{k}}^+\left[1+\cosh\left(\frac{\epsilon_{1,\bm{k}}^+}{T}\right)
\right]} + \frac{\sinh\left(\frac{\epsilon_{2,\bm{k}}^+}{T}\right)}
{\epsilon_{2,\bm{k}}^+\left[1+\cosh\left(\frac{\epsilon_{2,\bm{k}}^+}{T}\right)
\right]} 
\ee
and
\be
B_{\bm{k}} = \frac{\sinh\left(\frac{\epsilon_{1,\bm{k}}^+}{T}\right)}
{\epsilon_{1,\bm{k}}^+\left[1+\cosh\left(\frac{\epsilon_{1,\bm{k}}^+}{T}\right)
\right]} - \frac{\sinh\left(\frac{\epsilon_{2,\bm{k}}^+}{T}\right)}
{\epsilon_{2,\bm{k}}^+\left[1+\cosh\left(\frac{\epsilon_{2,\bm{k}}^+}{T}\right)
\right]}
\ee
The remaining notation is defined in \eqref{eq:HMF_matrix_elements}
and \eqref{eq:eigenenergies_terms}.

\section{\label{app:self_cons_AF}Antiferromagnetic mean-field self-consitency 
equations}

In Sec. \ref{subsec:mf_af} we extend the ``doped carrier'' MF theory to 
include the staggered magnetization decoupling channel and obtain
the MF Hamiltonian \eqref{eq:HMF_AF}.
In this paper we do not consider states with coexisting AF and SC order and,
in what follows, we set $b_0$ and $b_1$ to zero in \eqref{eq:HMF_AF}.
Since in this case spinons and dopons do not mix we can define two
spinon and two dopon bands, namely
\begin{gather}
\epsilon_{s,\bm{k}}^{\pm} = \pm\sqrt{\left(\alpha_{\bm{k}}^x\right)^2
+\left(\alpha_{\bm{k}}^z\right)^2+\nu_{\bm{k}}^2} \notag \\
\epsilon_{d,\bm{k}}^{\pm} = \left(1\mp2|m|\right) \gamma_{\bm{k}} - \mu_d
\label{eq:eigenenergies_AF}
\end{gather}
where $\nu_{\bm{k}} = -2 \left(J^*m-n\right)$.
In the absence of spinon-dopon hybridization $a_0 = 0$ and 
$\alpha_{\bm{k}}^z = -\left(\frac{3\tilde{J}}{4}\chi-\frac{t_1 x}{2}\right)
\left(\cos k_x + \cos k_y\right)$.

The resulting MF free-energy is
\begin{align}
\frac{F}{N} &= \frac{3\tilde{J}}{4}\left(\chi^2+\eta^2\right)+
2J^*m^2-4mn-\mu_d(1-x)- \notag \\
& \quad - \frac{T}{N} \sum_{\bm{k}}
\left\{ 
\ln \left[\cosh\left(\frac{\gamma_{\bm{k}}-\mu_d}{T}\right) +
\cosh\left(\frac{2m\gamma_{\bm{k}}}{T}\right)\right] +
\right. \notag \\
& \quad \quad \quad \quad \quad \quad
\left. + 
\ln\left[1+\cosh\left(\frac{\epsilon_{s,\bm{k}}^+}{T}\right)\right]
\right\}
\label{eq:free_energy_AF}
\end{align}
and the self-consistency equations
\be
\frac{\p F}{\p \chi} = \frac{\p F}{\p \Delta} = \frac{\p F}{\p m} =
\frac{\p F}{\p n} = \frac{\p F}{\p \mu_d} = 0
\ee
reduce to
\be
\chi = -\frac{1}{2N} \sum_{\bm{k}} \frac{\alpha_{\bm{k}}^z\left(
\cos k_x + \cos k_y\right)}{\epsilon_{s,\bm{k}}^+}
\frac{\sinh\left(\frac{\epsilon_{s,\bm{k}}^+}{T}\right)}{1+
\cosh\left(\frac{\epsilon_{s,\bm{k}}^+}{T}\right)} 
\ee
\be
\Delta = -\frac{1}{2N} \sum_{\bm{k}} \frac{\alpha_{\bm{k}}^x\left(
\cos k_x - \cos k_y\right)}{\epsilon_{s,\bm{k}}^+}
\frac{\sinh\left(\frac{\epsilon_{s,\bm{k}}^+}{T}\right)}{1+
\cosh\left(\frac{\epsilon_{s,\bm{k}}^+}{T}\right)} 
\ee
\be
m = \frac{1}{N} \sum_{\bm{k}} \frac{J^*m-n}{\epsilon_{s,\bm{k}}^+}
\frac{\sinh\left(\frac{\epsilon_{s,\bm{k}}^+}{T}\right)}{1+
\cosh\left(\frac{\epsilon_{s,\bm{k}}^+}{T}\right)} 
\ee
\be
n = -\frac{1}{2N} \sum_{\bm{k}} \frac{\gamma_{\bm{k}} 
\sinh\left(\frac{2m\gamma_{\bm{k}}}{T}\right)}
{\cosh\left(\frac{\gamma_{\bm{k}}-\mu_d}{T}\right) +
\cosh\left(\frac{2m\gamma_{\bm{k}}}{T}\right)} 
\ee
\be
x=1-\frac{1}{N} \sum_{\bm{k}} 
\frac{\sinh\left(\frac{\gamma_{\bm{k}}-\mu_d}{T}\right)}
{\cosh\left(\frac{\gamma_{\bm{k}}-\mu_d}{T}\right) +
\cosh\left(\frac{2m\gamma_{\bm{k}}}{T}\right)} 
\ee

\bibliography{hightc}

\begin{thebibliography}{92}
\bibitem[{POL()}]{POLARON}
\bibinfo{note}{Here we assume that the hole doped into the spin system does not
  create a ferromagnetic polaron. This is believed to be the case unless $J/t$
  is an order of magnitude smaller than observed in material compounds like the
  cuprates \cite{WA0111}}.
\bibitem[{BAN()}]{BAND_I}
\bibinfo{note}{Band insulators have two valence electrons per site, hence, a
  hole on a certain site $i$ leaves a charge $-e$ and spin $\pm \tfrac{1}{2}$
  in this same site}.
\bibitem[{ED()}]{ED}
\bibinfo{note}{In the electron doped regime the states
  $\ket{\up}_i$,$\ket{\down}_i$ and $\ket{0}_i$ rather represent a spin-up
  hole, a spin-down hole and a site with no holes respectively. Therefore, in
  that case, the operator in \eqref{eq:electron_dagger_enl} is a hole creation
  operator}.
\bibitem[{E_O()}]{E_OPERATOR}
\bibinfo{note}{In Sec. \ref{sec:mf} we introduce the fermionic representation
  of the lattice spin operators $\widetilde{\bm{S}}_i = \tfrac{1}{2} f_i\sd
  \bm{\sigma} f_i$, where $f_i\sd$ is the spinon creation spinor operator,
  which is valid upon the enforcement of the $f_i\sd f_i = 1$ constraint. In
  such a representation the operator in \eqref{eq:electron_dagger_enl} reduces
  to $ \tilde{c}_{i,\sigma}\sd = s_{\sigma} \tfrac{1}{\sqrt{2}}
  \widetilde{\mathcal{P}} ( f_{i,\sigma}\sd f_{i,\sigma} d_{i,-\sigma} -
  f_{i,\sigma}\sd f_{i,-\sigma} d_{i,\sigma}) \widetilde{\mathcal{P}} $. This
  expression, which holds as long as the local constraint $f_i\sd f_i = 1$ is
  enforced, can be modified so that it explicitly vanishes in case $f_i\sd f_i
  \neq 1$, specifically $ \tilde{c}_{i,\sigma}\sd = s_{\sigma}
  \tfrac{1}{\sqrt{2}} \widetilde{\mathcal{P}} (f_{i,-\sigma}f_{i,-\sigma}\sd
  f_{i,\sigma}\sd f_{i,\sigma}d_{i,-\sigma}- f_{i,\sigma}\sd f_{i,-\sigma}
  d_{i,\sigma} ) \widetilde{\mathcal{P}} $ (the overall $s_{\sigma}$ is missed
  in the similar expression presented in Ref. \onlinecite{RW0501})}.
\bibitem[{LAM()}]{LAMBDA}
\bibinfo{note}{The procedure to determine $\lambda$ differs from the one used
  in Ref. \onlinecite{RW0501} and so does the value of $\lambda$ itself. Note
  that, here, $\lambda$ follows from fitting numerical data at half-filling}.
\bibitem[{REG()}]{REGIONS}
\bibinfo{note}{Here we only consider the temperture range displayed in the MF
  phase diagrams in Ref. \onlinecite{RW0501}, in which case we always find that
  $\chi\neq0$. Only at higher temperatures does $\chi$ vanish and, thus, do the
  lattice spins decouple from each other at MF level}.
\bibitem[{PHA()}]{PHASES}
\bibinfo{note}{Similar phases also appear in the slave-boson MF phase diagram
  \cite{WL9603} since $\chi$ and $\Delta$ arise in both approaches upon the
  introduction of the fermionic spin representation and given that the
  spinon-dopon pairing transition which leads to non-zero $b_0$ and $b_1$ is
  analogous to the holon condensation transition in the slave-boson framework
  (we further elaborate on this topic in Appendix \ref{app:slave_boson})}.
\bibitem[{NER()}]{NERNST}
\bibinfo{note}{The spinon-dopon pairing transition is driven by the kinetic
  energy gain that results from overcoming the frustration due to the perfect
  N\'eel configuration of spins around the vacancy in the one-dopon state
  [Expression \eqref{eq:s_square}] which, in the above described MF theory, is
  assumed to be temperature independent. Since increasing temperature reduces
  the magnitude of the staggered moments around the vacancy, the stability of
  spinon-dopon mixing at high temperatures is overestimated. Hence, the large
  phase-space volume taken by the Nernst regime in Fig.
  \ref{fig:pdiagram_2}(a), which extends throughout most of the pseudogap
  region, should be reduced upon calculating \eqref{eq:s_square}
  self-consistently. Such a change could shrink the Nernst region for the
  parameter regime that concerns hole doped cuprates down to the smaller size
  observed by experiments \cite{OW0409}}.
\bibitem[{SLA()}]{SLAVE_BOSON}
\bibinfo{note}{The slave-boson formulation and its application to the $tt't''J$
  model have been extensively studied in the literature and we refer the reader
  to Refs.
  \onlinecite{B7675,C8435,BZ8773,BA8880,AZ8845,LN9221,WL9603,LN9803,LN0445} for
  additional information}.
\end{thebibliography}

\end{document}